\documentclass[a4paper,pra,aps,reprint,footinbib,superscriptaddress,showpacs]{revtex4-1}

\usepackage{amsmath,amsfonts,amssymb,bm}
\usepackage{dcolumn}
\usepackage[final]{graphicx}
\usepackage{bm}
\usepackage{comment}
\usepackage{color}
\usepackage{braket}
\usepackage{siunitx}
\usepackage[T1]{fontenc}
\usepackage[utf8]{inputenc}

\begin{document}

\bibliographystyle{apsrev4-1}

\title{Dynamics of parametric matter wave amplification}


\author{Robert B\"ucker}
\affiliation{Vienna Center for Quantum Science and Technology, Atominstitut, TU Wien, Stadionallee 2, 1020  Vienna, Austria}

\author{Ulrich Hohenester} 
\email{ulrich.hohenester@uni-graz.at}
\affiliation{Institut  f\"ur Physik, Karl--Franzens--Universit\"at Graz,  Universit\"atsplatz 5, 8010 Graz, Austria}

\author{Tarik Berrada}
\author{Sandrine~van~Frank}
\affiliation{Vienna Center for Quantum Science and Technology, Atominstitut, TU Wien, Stadionallee 2, 1020  Vienna, Austria}

\author{Aur\'{e}lien Perrin}
\affiliation{Vienna Center for Quantum Science and Technology, Atominstitut, TU Wien, Stadionallee 2, 1020  Vienna, Austria}
\affiliation{Laboratoire de physique des lasers, CNRS, Université Paris 13, 99 avenue J.-B. Clément, 93430 Villetaneuse, France}

\author{Stephanie Manz}
\affiliation{Vienna Center for Quantum Science and Technology, Atominstitut, TU Wien, Stadionallee 2, 1020  Vienna, Austria}
\affiliation{Center for Free-Electron Laser Science, Notkestrasse 85, 22607 Hamburg, Germany}
\affiliation{MPSD at the University of Hamburg, Notkestrasse 85, 22607 Hamburg, Germany}

\author{Thomas Betz}
\affiliation{Vienna Center for Quantum Science and Technology, Atominstitut, TU Wien, Stadionallee 2, 1020  Vienna, Austria}
\affiliation{Center for Free-Electron Laser Science, Notkestrasse 85, 22607 Hamburg, Germany}
\affiliation{Max-Planck-Institut für Kernphysik, Saupfercheckweg 1, 69117 Heidelberg, Germany}

\author{Julian Grond}
\affiliation{Institut  f\"ur Physik, Karl--Franzens--Universit\"at Graz,  Universit\"atsplatz 5, 8010 Graz, Austria}
\affiliation{Theoretische Chemie, Universit\"at Heidelberg, Im Neuenheimer Feld 229, 69120 Heidelberg, Germany}

\author{Thorsten Schumm}
\author{J\"org Schmiedmayer} 
\affiliation{Vienna Center for Quantum Science and Technology, Atominstitut, TU Wien, Stadionallee 2, 1020  Vienna, Austria}

\begin{abstract}

We develop a model for parametric amplification, based on a density matrix approach, which naturally accounts for the peculiarities arising for matter waves: 
significant depletion and explicit time-dependence of the source state population, long interaction times, and spatial dynamics of the amplified modes. 
We apply our model to explain the details in an experimental study on twin-atom beam emission from a one-dimensional degenerate Bose gas.

\end{abstract}

\maketitle


\section{Introduction}

Parametric amplification of quantum mode populations is a feature common to both light, and matter waves.
For light, optical parametric amplification~\cite{Burnham1970,Heidmann1987} is the key technique to populate twin-modes containing correlated photon pairs.
For atomic matter waves, numerous emission and amplification schemes have been demonstrated both in spontaneous~\cite{Chikkatur2000,Perrin2007,Jaskula2010a} and Bose-enhanced~\cite{Deng1999,Vogels2002,*Gemelke2005,*Campbell2006,Dall2009,*RuGway2011,buecker:11,Klempt2010,*Lucke2011,Bookjans2011,*Hamley2012,Gross2011} regimes.
Ideally, the populations of two modes, selected by fundamental conservation laws, grow identically during the amplification process, analogous to the signal and idler modes in parametric down-conversion of photons.
Key features of such twin-atom beams, such as suppressed relative and enhanced absolute number fluctuations~\cite{Jaskula2010a,buecker:11,Klempt2010,*Lucke2011,Bookjans2011,*Hamley2012,Gross2011}, non-trivial second-order correlations~\cite{Perrin2007,Dall2009,*RuGway2011} and quadrature squeezing~\cite{Klempt2010,*Lucke2011,Gross2011,Bookjans2011,*Hamley2012} have been shown experimentally.

A crucial difference between matter and photon twin-beams arises from the microscopic process driving the stimulated emission.
As photons do not interact, the amplification process has to be mediated by a medium which is being pumped by a strong light field.
Due to the relatively weak $\chi^{(2)}$ nonlinearity and the short, localized interaction in the medium, the conversion efficiency is low, allowing to neglect the depletion of the pump beam and to reduce the description to the signal and idler modes only (undepleted pump approximation).
For interacting matter waves, the pump field itself acts as the nonlinear medium.
Interaction times can be long, so that the depletion of the source state considerably affects the dynamics.
Furthermore, the coupling between source and amplified modes typically extends over the entire system size, and the mode structure can be more complex.
At finite temperatures, thermal phase fluctuations reduce the coherence, effectively depleting the source state~\cite{Perrin2012}.
Finally, the source population may explicitly depend on time, e.g. if the emission process starts before pumping is completed, the source is being replenished continuously, or other loss channels are present.

\begin{figure}[b]
\includegraphics{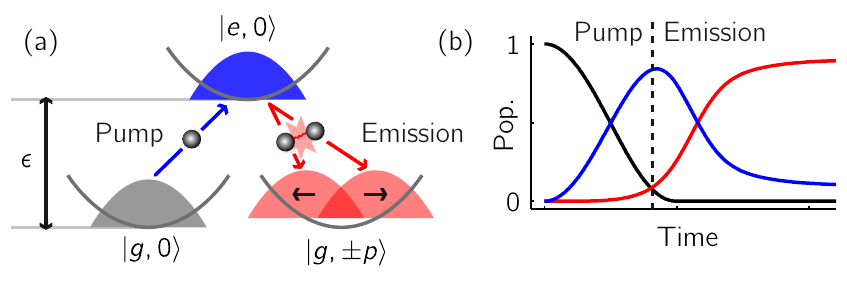}
\caption{(Color online) Overview of the twin-atom process. {(a)} Level scheme of the involved states and transitions.  {(b)} Qualitative population dynamics of the states as shown in (a) once the pumping process starts.}
\label{fig:overvew}
\end{figure}

In this article, we develop a simple and numerically tractable model for the description of stimulated matter wave emission into twin-beams.
Our approach is based on density matrices~\cite{vardi:01}, and extends previous theoretical studies which have concentrated on the regime of rapid scattering into many, weakly occupied modes, where the Bogoliubov approximation (equivalent to negligible depletion) holds~\cite{Pu2000,*Duan2000a,Bach2002,Zin2005,*Zin2006}.
In contrast to calculations based on the positive-P~\cite{Kheruntsyan2002,*Kheruntsyan2005a,*Kheruntsyan2005b,Deuar2007,Perrin2008,Carusotto2001} or truncated Wigner~\cite{Norrie2005,*Norrie2006} methods, our approach does not rely on stochastic sampling, and remains valid for long interaction times and arbitrary mode populations.
We will be primarily concerned with modeling of de-excitation experiments from a vibrational state~\cite{buecker:11}, where we record the population of amplified twin-beams over time, but our density matrix calculation can be applied to a much larger class of twin-beam experiments.

\section{Theory}

\begin{figure}
\includegraphics{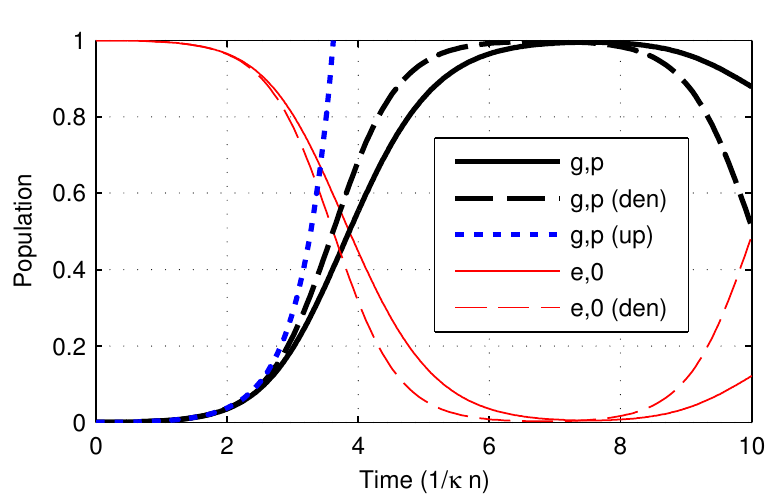}
\caption{(Color online) Population transients $\rho_{g,p}+\rho_{g,-p},\rho_{e,0}$ for the two-mode model of eq.~\eqref{eq:twomode}. 
In the simulations we use 700 atoms.
Thick and thin lines report the normalized population of $\ket{g,\pm p}$ and $\ket{e,0}$, respectively.
Solid and dashed lines correspond to simulations performed for the full Schr\"odinger equation and the density matrix approximation of eqs.~(\ref{eq:density}, \ref{eq:twoparticle}).
The dotted line indicates the growth of the emitted population for an undepleted pump.}
\label{fig:twomode}
\end{figure}

We consider a trapped, degenerate Bose gas in one dimension, such as the ground state $\ket{g,0}$ of a very elongated harmonic trap with weak confinement $\omega_x$ along its longitudinal direction $x$ and negligible occupation of transversally excited states.
Starting from $t=0$, we pump the system into a \emph{source} state $\ket{e,0}$, carrying a per-particle excess energy $\epsilon \gg \hbar \omega_x$ but leaving the spatial wave function along $x$ unchanged (fig.~\ref{fig:overvew}).
This energy can be stored in internal degrees of freedom~\cite{Pu2000,*Duan2000a}, dimer molecules~\cite{Kheruntsyan2002,*Kheruntsyan2005a,*Kheruntsyan2005b}, or in a vibrational state orthogonal to $x$~\cite{Spielman2006,buecker:11} as in the experiment described below.
Conservation laws for parity and momentum dictate that the state $\ket{e,0}$ can only decay via a two-body process into paired, propagating modes $\ket{g,\pm p}$ with identical population and opposite momenta centered around $\pm p \approx \pm \sqrt{2m\epsilon}$, where $m$ denotes the particle mass.
A slight shift may arise from mean-field effects~\cite{Krachmalnicoff2010}.
For a scheme populating twin-modes in a spinor gas~\cite{Klempt2010,*Lucke2011,Bookjans2011,*Hamley2012,Gross2011}, the equivalent states would be the $\ket{m=0}$ state as a source and two states with opposite magnetizations $\ket{\pm m}$ instead of $\ket{g,\pm p}$.

\subsection{Two-mode model}

If we consider a single pair of modes resonant with $\ket{e,0}$ and instantaneous excitation at $t=0$, the relevant part of the two-body interaction Hamiltonian can be written as
\begin{equation}\label{eq:twomode}
  \hat H_{\rm TM}=\kappa\left\lgroup\hat a_{g,p}^\dagger\hat a_{g,-p}^\dagger
  \left(\hat a_{e,0}^{\phantom\dagger}\right)^2+
  \left(\hat a_{e,0}^\dagger\right)^2
  \hat a_{g,p}^{\phantom\dagger}\hat a_{g,-p}^{\phantom\dagger}\right\rgroup\,,
\end{equation}
similar to parametric amplification in optics, where the strength of atom-atom interaction $\kappa$ corresponds to the nonlinear susceptibility.

A full numerical solution of the Schr\"odinger equation with the two-mode Hamiltonian \eqref{eq:twomode} can be  accomplished by expanding the many-body wave function in terms of bosonic Fock states, where $m$ atom pairs are promoted from $\ket{e,0}$ to $\ket{g,\pm p}$:
\begin{equation}\label{eq:psitwomode}
  \psi=\sum_{m=0}^{n/2}C_m
    \left(\hat a_{g,p}^\dagger\hat a_{g,-p}^\dagger\right)^m
    \left(\hat a_{e,0}^\dagger\right)^{n-2m}|0\rangle\,.
\end{equation}
Here $C_m$ are the wave function amplitudes in Fock space and $\ket{0}$ is the vacuum state.
The Schr\"odinger equation with the Hamiltonian $\hat{H}_{\rm TM}$ is solved with the Crank-Nicolson method according to the prescription given in ref.~\cite{javanainen:99}.

Solid lines in fig.~\ref{fig:twomode} show results for a simulation of $n=700$ atoms, which initially all reside in the transversally excited $\ket{e,0}$ state.  
The non-linear interaction of eq.~\eqref{eq:twomode} promotes atoms pairwise to the twin-atom states $\ket{g,\pm p}$.  
Initially the process is slow and governed by spontaneous scatterings.  
Only when a sufficient population has built up in $\ket{g,\pm p}$, say at times around $2/\kappa n$, the scattered atoms act as a seed for the ensuing rapid stimulated emission which continues until the $\ket{e,0}$ reservoir is emptied.  
Finally, owing to the second term in parentheses of eq.~\eqref{eq:twomode}, atoms scatter back from the twin-atom states to $\ket{e,0}$.

Unfortunately, for many near-resonant modes, the exponentially increasing size of the Hilbert space prohibits a generalization of this direct approach.
To overcome this problem, we introduce a density-matrix description for the two-level problem, which we will extend to many modes in the next section.
The lowest moments are the densities $\rho_{e,0}=\langle\hat a_{e,0}^\dagger\hat a_{e,0}^{\phantom\dagger}\rangle$ and $\rho_{g,p}=\langle\hat a_{g,p}^\dagger\hat a_{g,p}^{\phantom\dagger}\rangle$.
Their dynamic equations can be obtained from the Heisenberg equations of motion, 
\begin{equation}\label{eq:density}
  \dot\rho_{g,\pm p}=-\frac 12\dot\rho_{e,0}=2\kappa\,\Im (\Delta)\,,\quad
  \Delta=\langle\hat a_{g,p}^\dagger\hat a_{g,-p}^\dagger
  \bigl(\hat a_{e,0}^{\phantom\dagger}\bigr)^2\rangle\,.
\end{equation}
Here, the densities are driven by the two-particle coherence $\Delta$, whose time evolution is in turn governed by three-particle couplings.
To truncate this hierarchy of equations of motion, we introduce a correlation expansion in the spirit of ref.~\cite{vardi:01}, and factorize all three-particle density matrices into densities of lower order, e.g. through $\langle\hat a_{g,p}^\dagger\hat a_{g,-p}^\dagger\hat a_{e,0}^\dagger \hat a_{g,p}^{\phantom\dagger}\hat a_{g,-p}^{\phantom\dagger} \hat a_{e,0}^{\phantom\dagger}\rangle\approx \rho_{g,p}\rho_{g,-p}\rho_{e,0}$.
This finally yields
\begin{equation}\label{eq:twoparticle}
\dot \Delta \approx i \kappa \left[ \rho_{e,0} (\rho_{e,0} - 1) (2\rho_{g,p} + 1)
- 2\rho_{g,p}^2 (2\rho_{e,0} + 1) \right].
\end{equation}
Initially, $\rho_{g,p} = 0$ and $\rho_{e,0} = n\gg 1$, and eq.~(\ref{eq:twoparticle}) reduces to $\dot\Delta\approx i\kappa n^2$.
Consequently, the emitted population grows quadratically as $\rho_{g,\pm p} \approx (\kappa n t)^2$, which for short times $t$ is consistent with the result for an undepleted pump~\cite{Bach2002,Zin2005,*Zin2006}: $\rho^\text{(up)}_{g,\pm p}=\sinh^2(\kappa n t)$.
In eq.~\eqref{eq:twoparticle}, the second term in brackets accounts for the backscattering of population into the source state.
A comparison of the results of the density matrix equations~(\ref{eq:density},~\ref{eq:twoparticle}), see dashed lines in fig.~\ref{fig:twomode}, with the exact solution shows good agreement.


\subsection{Multi-mode model}

In a realistic matter-wave amplifier, several twin-modes become populated simultaneously, as the amplification bandwidth is broadened by the mean field of the source.
In contrast to a full wave function approach, it is numerically straightforward to extend the density matrix framework of eqs.~(\ref{eq:density},\ref{eq:twoparticle}) to a multi-mode Hamiltonian

\begin{equation}\label{eq:multimode}
  \hat H_{\rm MM}=
  \frac 12\sum_{ij}\left\lgroup\kappa_{ij}
  \hat a_{g,i}^\dagger\hat a_{g,j}^\dagger
  \left(\hat a_{e,0}^{\phantom\dagger}\right)^2+\mbox{H.c.}\right\rgroup\,,
\end{equation}
where $\kappa_{ij}$ is the interaction matrix element between the source and the different spatial modes $\ket{g,i}$.
Momentum and parity conservation are fulfilled due to the symmetries of the matrix elements $\kappa_{ij}$.
The number of necessary modes $M$ can be estimated from the per-particle mean field energy $\mu$ of  the source~\cite{Bach2002}: $M \sim \mu/\hbar\omega$, where $\hbar\omega$ is the typical energy level spacing of adjacent modes.
The multi-mode description inherently includes spatial dynamics, such as the propagation of twin-beam wave packets.

As for the two-mode case outlined in the previous section, we employ a factorization scheme for many-particle density matrices to obtain the density-matrix equations for the many-mode Hamiltonian~\eqref{eq:multimode}.
Let $\rho_{g,ij}$ be the single-particle reduced density matrix for the twin-atom states, generalizing $\rho_{g,\pm p}$ of the two-mode case.
The dynamic equations for $\rho$ directly follow from the Heisenberg equations of motion
\begin{subequations}
\begin{eqnarray}
  \dot\rho_{e,0} &=& -2\Im\bigl( \sum_{ij}\kappa_{ij}\Delta_{ij}\bigr)\\
  \dot\rho_{g,ij} &=& \sum_k\left(\kappa_{ik}\Delta_{kj}-\Delta_{ik}^*\kappa_{kj}\right).
\end{eqnarray}
\end{subequations}
Here $\Delta_{ij}=\langle\hat a_i^\dagger\hat a_j^\dagger(\hat a_0)^2\rangle$ is the two-particle coherence between source and emitted modes, whose time evolution can be evaluated to
\begin{eqnarray}
  && i\dot\Delta_{ij} = -\kappa_{ij}\rho_{e,0}(\rho_{e,0}-1)+
  (2\rho_{e,0}+1)\sum_{kl}\kappa_{kl}\rho_{g,ik}\rho_{g,jl}\nonumber\\&&\qquad
  -\rho_{e,0}(\rho_{e,0}-1)\sum_k\left(\kappa_{ik}\rho_{g,kj}+\kappa_{jk}\rho_{g,ki}\right)\,.
\end{eqnarray}
With this result, the dynamics of the twin-mode populations can be derived, similarly to the two-mode case.
%

\section{Experiment}

\begin{figure}
\includegraphics{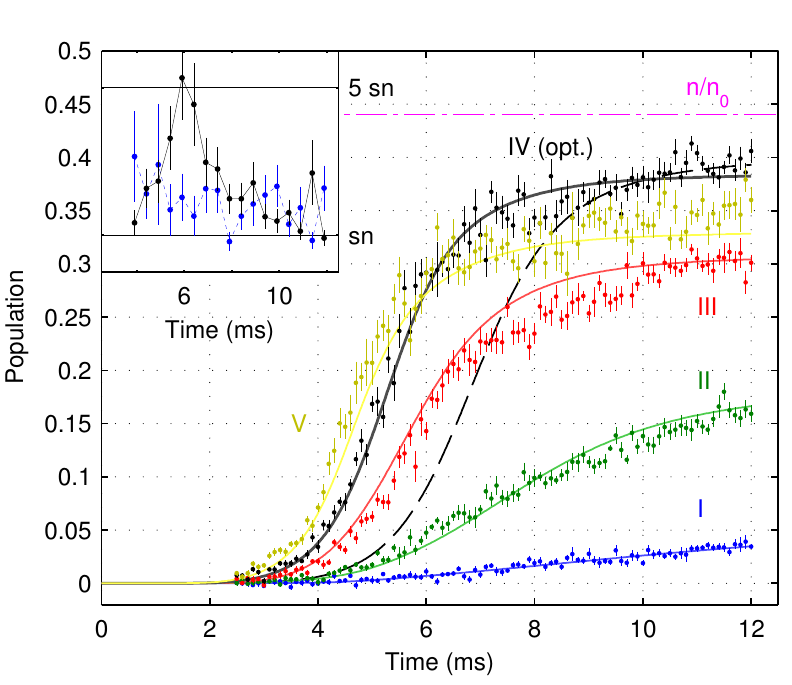}
\caption{
(Color online) Population of twin-atom modes, normalized to the total atom number.
The data points with error bars report the experimental results for optimal (IV) and scaled excitation (other lines) with scaling factors of approximately 0.3 (I), 0.5 (II), 0.8 (III), and 1.2 (V).
Solid lines: corresponding theory results $\sum_{k} \rho_{g,kk}$.
Dashed line: theory for optimal excitation, where additional spontaneous emission processes are neglected.
Dash-dotted line: Upper bound imposed by the population of the Onsager-Penrose mode.
(Inset) Variance of emitted fraction, relative to shot noise (sn), for optimal excitation (solid, black) and scaling 0.3 (dotted, blue). Each point is an average over seven adjacent times and corrected for imaging noise~\cite{Buecker2009} and total atom number fluctuations.
}
\label{fig:pop}
\end{figure}

As a representative example of twin-beam creation, in this paper we consider a collisional de-excitation scheme based on a one-dimensional, quantum degenerate Bose gas of Rubidium-87 atoms trapped on a chip~\cite{Reichel2011,Trinker2008a}, as described in more detail in ref.~\cite{buecker:11}.  
Within \SI{5}{ms}, atoms are pumped into the source state $\ket{e,0}$, which is a vibrationally excited transversal state of the confining waveguide potential.
This pumping is accomplished via a mechanical optimal control protocol, populating $\ket{e,0}$ with almost unity efficiency by non-adiabatic translation of a strongly anharmonic trapping potential along the transversal direction $y$ (see appendix~\ref{sec:trap}).
In fig.~\ref{fig:pop} we show the population of the twin-beam modes, relative to the total atom number $n\approx 800$, as a function of time after starting the excitation sequence.
Experimental points are obtained by counting photons in appropriately defined regions within fluorescence images taken after \SI{46}{ms} of free expansion, see appendix~\ref{sec:detect} for details.
One observes that for the optimal ramp (black markers, series IV) the twin-atom population increases over approximately \SI{10}{ms} and finally reaches a plateau value.
About 40~percent of the atoms are emitted into twin beams.
We repeated the measurement several times, scaling the amplitude of the optimal trap motion used for pumping by different factors \footnote{Note that this procedure is not equivalent to using numerically optimized excitations for different efficiencies, and may lead e.g. to increased collective oscillations after the excitation, which is irrelevant for the emission dynamics.}.
For these non-optimal excitation protocols (series I-III and V) the final twin-atom populations are reduced.

\begin{figure}[b]
\includegraphics{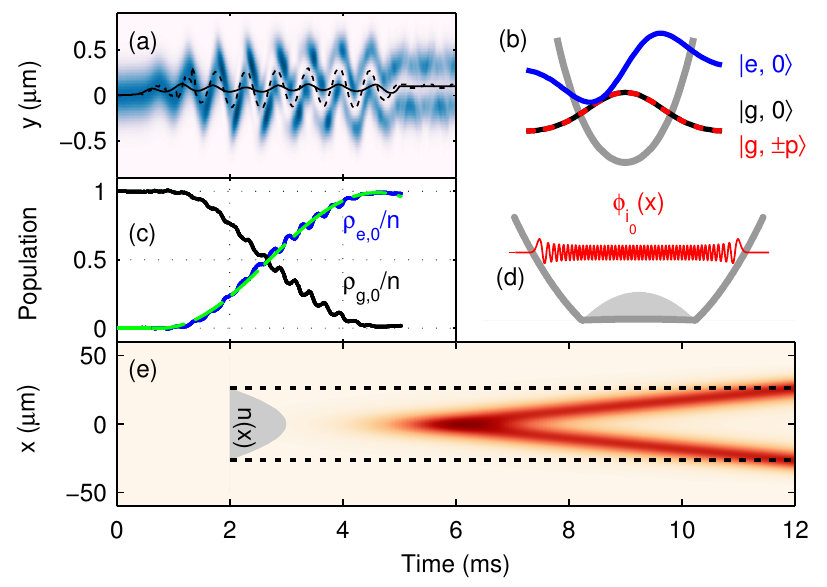}
\caption{(Color online) Numerical twin-beam dynamics results, from the simulations for our experiment.
{(a)} Density distribution along the transversal direction $y$.
The solid line indicates the driven motion of the trap minimum.
The dashed line shows the collective oscillation frame used to define the vibrational states.
{(b)} Wave functions of the vibrational states along $y$, corresponding to $\ket{g,0},\ket{g,\pm p}$ and $\ket{e,0}$, respectively, at zero collective displacement.
{(c)} Populations of $\ket{g,0}$ and $\ket{e,0}$, derived by projecting the numerical results shown in (a) and neglecting the twin-beam emission.
The dashed (green) line shows a constant Rabi coupling model.
{(d)} Illustration of one particular twin-beam mode $\phi_{i_0}$ with energy $\epsilon$ within the effective longitudinal potential, which is defined by the trap confinement and the mean field of the source cloud.
{(e)} Spatial evolution of the twin-beam modes along the longitudinal axis, which are created in the center of the condensate region and then propagate out.
The shaded area corresponds to the density distribution of the initial state, the dashed lines indicate its radius. 
}
\label{fig:shaking}
\end{figure}

To model these experiments, we need to describe the time-dependent excitation (pumping) of atoms into the source state with energy $\epsilon$, see fig.~\ref{fig:overvew}(b), taking place on a time scale on the order of $\sim 10 h/\epsilon \approx \SI{5}{ms}$ with $h$ the Planck constant, which is similar to that of the emission.
Additionally, we have to account for the thermal depletion of the source mode and the spatial dynamics of the twin beams.  
As will be demonstrated next, all of this can be conveniently accomplished within the density matrix approach.

\subsection{Pumping}

Fig.~\ref{fig:shaking}(a) shows the time-dependent position of the trap minimum (solid line) together with the time-dependent density distribution along $y$, which is obtained from the solution of the one-dimensional Gross-Pitaevskii equation (GPE) along $y$.
We can approximate the excitation dynamics by a two-mode description comprising the states $\ket{g,0}$ and $\ket{e,0}$, which are chosen as displaced eigenstates~\cite{DeOliveira1990} of the harmonic part of the waveguide potential with a common, periodic displacement (see panel (b)~).
This displacement is reflecting the collective oscillation (fig.~\ref{fig:shaking}(a), dashed line), which is necessary to drive the system into the strongly anharmonic part of the potential where the state transfer occurs, but is otherwise irrelevant for the dynamics between the states.
Projecting the GPE results onto the displaced state basis yields the population of $\ket{g,0}$ and $\ket{e,0}$, as depicted in panel (c).
A near-resonant Rabi coupling with an initial time delay (dashed line in panel (c)~) can be fitted to the population dynamics to provide a suitable model for the following calculation steps.
The optimal control sequence (leading to series IV in fig.~\ref{fig:pop}) then corresponds to a $\pi$-pulse which completely inverts the system from $\ket{g,0}$ to $\ket{e,0}$. 
See appendix~\ref{sec:transversal} for a more comprehensive discussion of the transversal dynamics.

\subsection{Thermal source depletion}

We now turn to the emission dynamics along the longitudinal direction $x$.
Especially for an elongated system, as studied here, thermal longitudinal phase fluctuations are expected to have a strong influence on the experimental results by effectively depleting the source mode~\cite{Perrin2012}.
We assume that the atoms are initially in the Thomas-Fermi ground state of a harmonic trap along $x$, and account for phase fluctuations through a density matrix $\rho(x,x')$ obtained from quasi-condensate theory~\cite{petrov:00}, where the Bogoliubov excitations are populated thermally.
Next, we split $\rho$ into a condensate (Onsager-Penrose mode) and a thermally excited part $\rho_{\rm th}(x,x')$,
\begin{equation}\label{eq:condensate}
  \rho(x,x')=n_0\phi_0^*(x)\phi_0(x')+\rho_{\rm th}(x,x')\,,
\end{equation}
where $n_0$ is the largest eigenvalue of the density matrix and $\phi_0(x)$ the corresponding eigenfunction.
For our calculation we assume $T=\SI{25}{nK}$, which is compatible with experimental observations~\footnote{Note that the non-condensed modes, described by $\rho_{th}(x,x')$ in eq.~(\ref{eq:condensate}), rapidly re-thermalize~\cite{Mazets2011}, impeding an unambiguous determination of $T$ from experimental data.
The temperature-dependent emission and thermalization dynamics will be subject of a future publication.}
and leads to $n_0\approx 0.44\,n$ with $n\approx 800$ (see dash-dotted line in fig.~\ref{fig:pop}).
As the twin-atom production is driven by a fixed phase relation between the source and twin-atom states, eq.~\eqref{eq:density}, the non-condensed part is expected to have only little influence on the emission dynamics, and will be neglected in the following.

\subsection{Twin-beam states}

To apply the multi-mode description of eq.~\eqref{eq:multimode} to our finite-size system, we compute a set of $\approx 30$ highly excited (real) single-particle states $\phi_i(x)$ at energies around $\epsilon = \hbar^2 k_0^2 / 2m$, for an effective potential including the mean field of the source state. 
One typical state is depicted in fig.~\ref{fig:shaking}(d).
The coupling matrix elements of $\hat H_{\rm MM}$ can now be expressed as 

\begin{align}
\kappa_{ij} = g &\int \left| \psi_g(z) \right|^4 \mathrm{d}z 
\int \psi^*_g(y)^2\psi_e(y)^2\mathrm{d}y \\
&\times \int \phi^*_i(x) \phi^*_j(x)\phi_0(x)^2\mathrm{d}x,  \nonumber
\end{align}
with $\psi_{e,g}(y)$ and $\psi_g(z)$ denoting the wave functions of the transversal states defined above, and $g=4\pi \hbar^2 a_s / m$, where $a_s$ and $m$ are scattering length and mass of the atoms.
The full dynamics of the twin-atom production is now governed by the Hamiltonian
\begin{equation}\label{eq:twinproduction}
  \hat H=\hat H_0+\hat H_{\rm pump}+\hat H_{\rm MM},
\end{equation}
where $\hat H_0$ accounts for the free evolution of $\ket{e,0}$, $\ket{g,0}$, and $\ket{g,i}$, and $\hat H_{\rm pump}$ describes the pumping process through the Rabi-type excitation described above.

\subsection{Results: optimal excitation}

For the Hamiltonian in eq.~\eqref{eq:twinproduction}, we numerically solve the equations of motion of the density matrices.
Fig.~\ref{fig:shaking}(e) shows the computed real-space density of twin-atoms as a function of time.
One observes that the twin-atoms are emitted in the condensate region initially, and then propagate as packets with group velocities $\pm \hbar k_0/m$, reaching the edges of the Thomas-Fermi distribution a few \si{ms} after the end of the excitation pulse.
All parameters of our model are obtained for realistic trap potentials (source excess energy $\epsilon\approx h\times\SI{1.8}{kHz}$) and experimental parameters taken from ref.~\cite{buecker:11}.
For the optimal excitation protocol, the slope and final value of the emission dynamics (dashed line in fig.~\ref{fig:pop}(a) ) is in good agreement with the experimental observation (black points), with the exception of a time delay of $\approx\SI{2}{ms}$.
This indicates that we underestimate the emission rate at early times, where even a very small twin-beam population suffices to trigger Bose-enhanced emission.
Such scattering may be caused by the non-condensate source modes, or a non-thermal population of high-momentum modes due to technical noise causing premature excitation.
To account for this, we add a weak channel for scattering atoms into longitudinal modes with momenta around $\pm k_0$ which act as an additional seed in excess of vacuum fluctuations.
Good agreement can be reached for a scattering rate of  $\Gamma \approx \SI{0.4}{s^{-1}} \cdot n_e^2$, which is equivalent to typically $\sim 3\cdot10^{-2} \kappa_{ii} \cdot n_e^2$ per mode, with $n_e = \langle\hat a_{e,0}^\dagger \hat a_{e,0}\rangle$.
Indeed, with this addition the simulation (series IV, solid line) accurately reproduces the experimental result. 

\subsection{Results: modified excitation}

We next address the results for scaling the amplitude of the trap motion by different factors (series I-III, V), which again can be well described by Rabi pulses of different area.
The couplings used in the calculations for the twin-beam population growth are derived from observation of the transversal momentum distribution during the excitation~(see appendix~\ref{sec:transversal}).
For series I, IV, and V in fig.~\ref{fig:pop}, this procedure yields good agreement without any further free parameters. 
Only for intermediate excitation scalings (series II, III), a slight adjustment of the excitation coupling was necessary, which we attribute to the approximations made in the pumping model (see above).
Note, that for the lowest scaling Bose enhancement is weak, which provides a stringent means for determining $\Gamma$.
The complete set of experimental results can be nicely modeled by our theory (solid lines in fig.~\ref{fig:pop}).

To illustrate the amplified character of the twin-beam creation, we investigate the fluctuations of the twin beams population $N$ at different times.
In fig.~\ref{fig:pop} (inset) we show the measured relative variances $\xi^2 \propto \sigma^2_N/\bar{N}$ of the relative population of emitted pairs over many experimental realizations.
They are corrected for imaging noise and total atom number fluctuations, and normalized to an approximation to the shot noise expected for random spontaneous emission, neglecting pump depletion and temperature fluctuations.
See appendix \ref{sec:variance} for details on the variance calculation.
For optimal excitation (black dots, solid line), a pronounced peak near the maximum slope of the population growth indicates the exponential amplification of initial fluctuations.
Such behavior is absent in the experiment with the weakest excitation (blue dots, dashed line).
In contrast, the relative number fluctuations of the twin-atom clouds are strongly suppressed, as discussed in detail in ref.~\cite{buecker:11}.

\section{Conclusion}

In conclusion, we have derived a density matrix approach to quantitatively analyze emission of matter waves in the strongly Bose-enhanced regime, which has recently been reached in various experiments. 
Neither the Bogoliubov approximation, nor stochastic methods are employed, making our approach eligible for strong depletion of the source and long interaction times.
We have presented experimental results for twin-beam population growth in a one-dimensional degenerate Bose gas, which are governed by source depletion, spatial dynamics, and explicit time-dependence of the source population.
The good agreement between experimental results and theory predictions suggest that the physics underlying amplified emission of twin-atoms in a real experiment is captured by our model.
Next steps will comprise further experimental and theoretical studies on properties of strongly populated twin-atom beams beyond single-particle densities, with particular regard to second-order correlations, revealing number squeezing~\cite{Jaskula2010a,buecker:11,Klempt2010,*Lucke2011,Bookjans2011,*Hamley2012,Gross2011} or violation of classical inequalities~\cite{Kheruntsyan2012}.

We acknowledge support from the Austrian Science Fund projects QuDeGPM (EuroQUASAR), CAP and SFB FoQuS, the FWF doctoral programme CoQuS (W 1210), the EU project AQUTE, and the Humboldt-Stiftung. We wish to thank I.~Bouchoule, K.~Kheruntsyan, I.~Mazets, H.~Ritsch, and J.-F.~Schaff for helpful discussions.

\appendix

\section{Trap preparation}
\label{sec:trap}

The ultracold Rubidum-87 gas acting as matter-wave source is prepared using laser cooling and then loaded into an atom chip wire trap~\cite{Trinker2008a,Reichel2011}, where forced evaporative cooling to quantum degeneracy occurs.
The initially transversally symmetric Ioffe-Pritchard field configuration created by the chip wires is modified by radio-frequency dressing \cite{Lesanovsky2006,Schumm2005a}.
Typically used for creating double well potentials, this technique also allows for the introduction of anharmonicity and anisotropy to a single trap when the dressing strength is kept below the point where actual splitting of the potential occurs.
Using chip wires running parallel to the main trapping wire, we apply an ac magnetic field of $\sim\SI{0.75}{G}$ peak-to-peak amplitude at a frequency red-detuned by $-\SI{54}{kHz}$ with respect to the atomic Larmor frequency near the trap minimum (\SI{824}{kHz}).

The resulting potential can be calculated numerically by means of a Floquet analysis \cite{Shirley1965}.
In the two transversal directions it can be approximated by quartic polynomials of the form $V/\hbar=\frac{1}{2} \omega (r/r_0)^2 + \lambda (r/r_0)^4$, where $r_0 = \sqrt{\hbar/m\omega}$ (Duffing oscillator) with the atomic mass $m$. 
For the $y$-direction, along which the excitation is performed the parameters are $\omega_y/2\pi=\SI{1.64}{kHz}$, $r_{0,y}=\SI{0.266}{\micro m}$ and $\lambda_y/2\pi = \SI{74.9}{Hz}$.
For the $z$-direction perpendicular to the excitation motion the parameters are $\omega_z/2\pi=\SI{2.50}{kHz}$, $r_{0,z}=\SI{0.216}{\micro m}$ and $\lambda_z/2\pi=\SI{25.9}{Hz}$.
By solving the Schrödinger equation for these potentials we can determine the first trap levels along $y$ and $z$ as $\left[E_{y,1}, E_{y,2}, E_{z,1},E_{z,2}\right]/2\pi\hbar = \left[1.82, 3.78, 2.56, 5.18\right]~\si{kHz}$, if the respective zero-point energies are subtracted.
Along the longitudinal $x$-axis, the harmonic trap frequency $\omega_x/2\pi=\SI{16.3}{Hz}$ is determined by observation of a deliberately excited sloshing mode of the atom cloud.

The fast transversal motion of the potential is accomplished by applying a modulated current to an auxiliary wire running parallel to the main trapping wire, moving the trap minimum by \SI{26}{nm/mA} along $y$ and \SI{9}{nm/mA} along $z$.
Due to the anisotropy of the transversal potential, the motion along $z$ is off-resonant and has no significant influence.

\begin{figure}

\centering
\includegraphics[width=86mm]{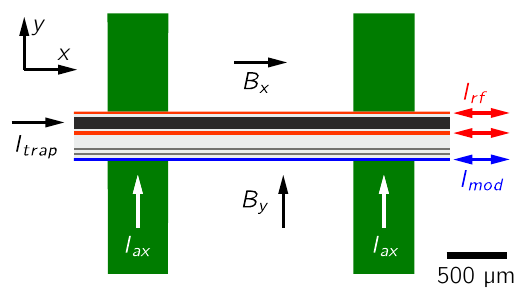}

\caption{(Color online) Schematic of the atom chip layout.  
The waveguide potential is formed by the current $I_\mathrm{trap}$ through the main trapping wire and a static magnetic field $B_{y}$. 
On a separate chip layer, currents $I_\mathrm{ax}$ in broad wires provide axial confinement. 
An external field $B_{x}$ completes the Ioffe-Pritchard configuration. 
The radio frequency dressing currents $I_\mathrm{rf}$ are applied to wires in parallel to the trapping wire. 
Finally, the modulation of the trap position is accomplished by a current $I_\mathrm{mod}$ in an auxiliary wire.}
\label{fig:wires}

\end{figure}

The geometry of all involved chip wires and homogeneous offset fields is shown in fig.~\ref{fig:wires}.


\section{Detection and image analysis}
\label{sec:detect}

\begin{figure}
\includegraphics{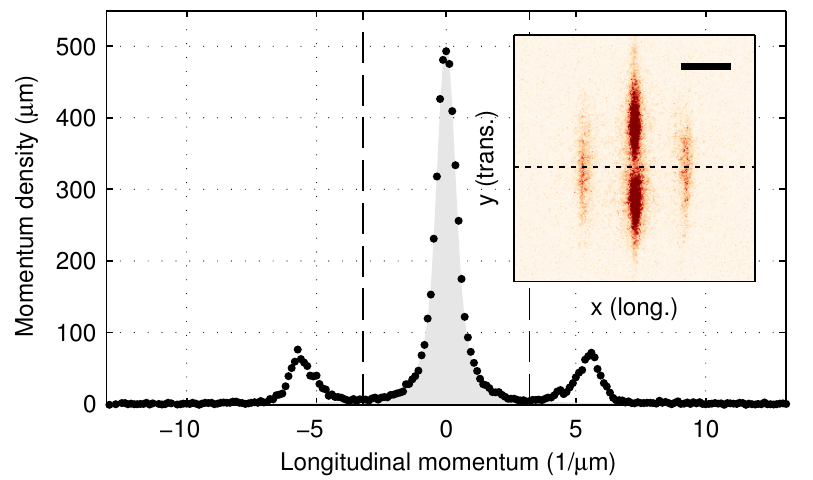}
\caption{(Color online) Experimental image analysis. {(Inset)} Typical experimental image data for optimal excitation and $t\leq\SI{5.5}{ms}$, averaged over 12 shots.
As the image is taken after \SI{46}{ms} of expansion time, it predominantly reflects the initial momentum distribution.
The scale bar corresponds to a spatial distance of \SI{187}{\micro m}, equivalent to the typical momentum of $\SI{5.5}{\micro m^{-1}}\approx k_0$.
The dashed line indicates the transversal center-of-mass position of the emitted clouds.
{(Main plot)} Longitudinal momentum distribution obtained from integration over the image. 
The shaded area indicates a fitted thermal distribution with temperature $T\approx\SI{25}{nK}$.
Dashed lines: range (source region) outside of which atoms are counted as emitted population.}
\label{fig:image}
\end{figure}

After starting the excitation sequence, we run the excitation and emission process for a given time $t$.
At this time, we suddenly switch off the trapping potential, implying that for $t<\SI{5}{ms}$ the excitation process is still incomplete.
The fast transversal expansion of the cloud due to the tight waveguide confinement causes atom interactions to vanish rapidly, and the ensuing expansion can be considered ballistic.
After $t_\mathrm{tof} = \SI{46}{ms}$ of expansion, we take a fluorescence image~\cite{Buecker2009}, fully integrating over the $z$-direction~(see inset of fig.~\ref{fig:image}).
Along $y$ (i.e., integrating over $x$), the resulting image represents the initial momentum distribution, as the transversal cloud size before expansion is negligible (far field).
If we express momenta as wave numbers $k_y$, a distance $\delta y$ in the image hence corresponds to $\delta k_y = \alpha \, \delta y$ with $\alpha = m/\hbar t_\textrm{tof} \approx \SI{0.03}{\micro m^{-2}}$.
Along the longitudinal direction $x$, the far field condition is not fully reached due to the initial condensate radius of $L\approx\SI{20}{\micro m}$ and the typical momenta in the quasi condensate as given by the thermal phase coherence length $l_\phi$~\cite{petrov:00}: $k_\phi =l_\phi^{-1}\sim\SI{0.1}{\micro m^{-1}}\approx\alpha L$.
Using an appropriate model for the initial density and momentum distributions (classical fields method~\cite{Jacqmin2012}), we can fit a longitudinal thermal distribution to the expanding cloud (fig.~\ref{fig:image}), yielding an estimate to its temperature. 
Within our current analysis, this relies on the assumption that the excitation and emission process does not lead to a strong modification of the momentum spectrum at short times $t$.
A more detailed study of the interplay between thermal effects and the emission dynamics is under way.

On the other hand, the center momentum of the emitted atoms is $k_0\approx\SI{5.5}{\micro m^{-1}} \gg (\alpha L, k_\phi)$, hence the longitudinal overlap of the source cloud and the emitted beams is negligible.
The fraction of emitted atoms at time $t$ (as shown in fig.~\ref{fig:pop}) can thus be determined by simple counting of fluorescence photons in appropriately defined ranges of the longitudinal distribution (outside of dashed lines in filg.~\ref{fig:image}).
For each combination of $t$ and excitation amplitude we repeated the experiment typically 12 times to obtain reliable estimates for the mean and variance of the emitted fraction.

\section{Transversal dynamics}
\label{sec:transversal}

\begin{figure*}
\begin{minipage}{\textwidth}
\includegraphics{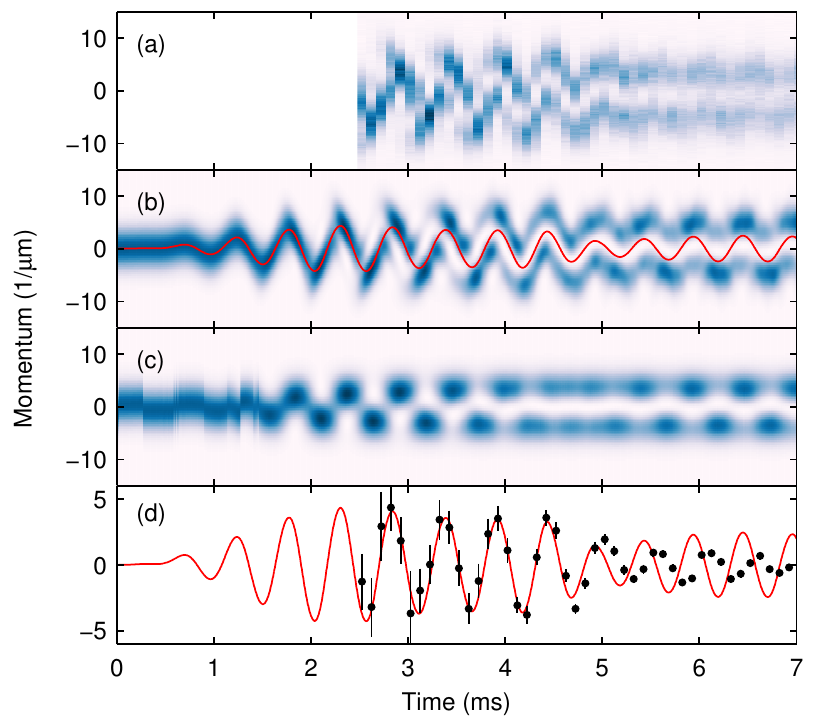}
\includegraphics{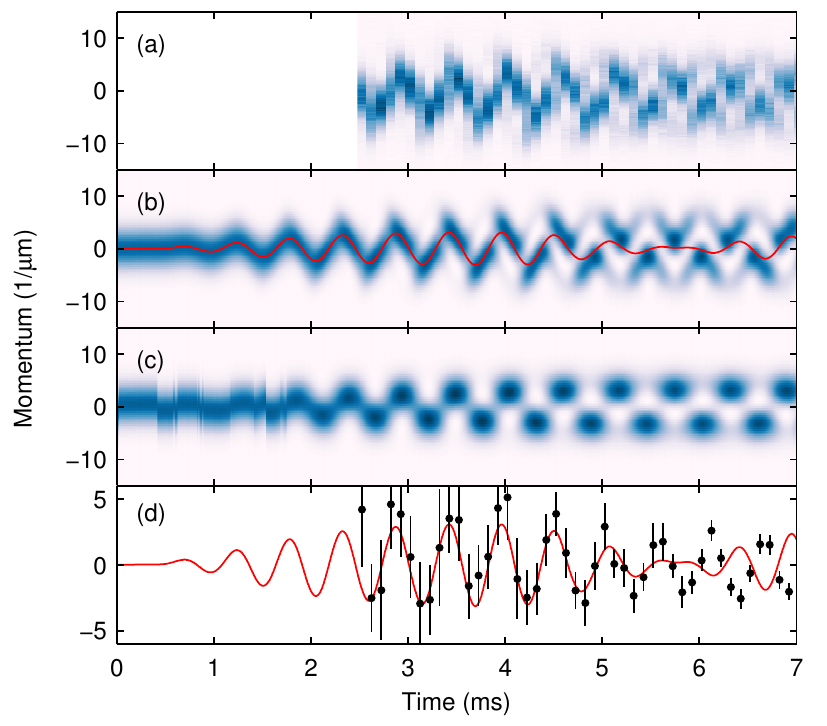}
\end{minipage}

\caption{(Color online) Transversal momentum dynamics. Left: optimal excitation, right: excitation amplitude scaled by $\approx 50~\%$. 
(a) Transversal momentum distribution of the excited cloud, obtained from fluorescence images by integrating over $x$ within the source region (dashed lines in fig.~\ref{fig:image}).
(b) Numerical result obtained from one-dimensional GPE.
For the scaled ramp, the scaling amplitude is fitted, whereas there are no free parameters for the optimal ramp.
Solid line: numerically optimized co-oscillating reference frame position $\tilde{k}_y(t)$ as described in the text.
(c) Momentum distribution obtained from the two-state approximation within the co-oscillating frame.
This result corresponds to the effective populations used for the emission calculation~(fig.~\ref{fig:exc}a).
(d) Dots: transversal center-of-mass momentum of the emitted atoms, as extracted from images (dashed line in fig.~\ref{fig:image} (inset) ). 
Solid line: same as in (b).}
\label{fig:transdyn}
\end{figure*}

%

The vibrationally excited state~$\ket{e,0}$ of the anharmonic transversal potential is being populated by oscillatory displacement of its origin $V(y-y_0(t))$, following an optimal control ramp.
To determine the function $y_0(t)$, we propagate the wave function with the one-dimensional Gross-Pitaevskii equation (GPE), starting from its ground state, and maximize the overlap with the first excited state at $t=\SI{5}{ms}$.
At $t>\SI{5}{ms}$, excitation stops and $y_0$ is held constant.
As outlined above, our experimental time-of-flight images, integrated over $x$ within the bounds shown as dashed lines in fig.~\ref{fig:image}, directly reflect the momentum density distribution along $y$, allowing direct comparison to the results numerically obtained from GPE (fig.~\ref{fig:transdyn}a,b).
Excellent agreement can be reached, if an independently measured linear response function of the electronics employed for the experiments is taken into account, which slightly modifies the $y_0(t)$ trajectory in the experiment.
Once the fraction of emitted twin-beams becomes significant ($t\gtrsim\SI{5}{ms}$), the agreement gradually gets worse, as damping due to the emission is not accounted for in the excitation model.
For the scaled control ramps, we left the amplitude scaling as free parameter in the GPE simulation and directly optimized agreement with the experimentally observed dynamics.

As described in the main text, to properly define the orbitals of the states $\ket{g,0}$ and $\ket{e,0}$ we transfer into a coordinate system that is co-oscillating with the (quasi-classical) sloshing of the cloud within the harmonic part of the potential $V(y)$.
This collective oscillation is inherent to our excitation technique, as it is necessary to access the anharmonic part of the potential. 
Within the co-oscillating frame, the anharmonicity acts as an oscillating force in excess of the restoring force, enabling the transfer into an excited eigenstate~\cite{Jirari2009,Khani2009}.
Consequently, we approximate the wave functions of $\ket{g,0}$ and $\ket{e,0}$ with the displaced first and second eigenfunctions of a harmonic oscillator~\cite{DeOliveira1990}: 

\begin{align}
\ket{g,0} &= \hat{D}[\alpha(t)] \ket{0} \nonumber \\
\ket{e,0} &= \hat{D}[\alpha(t)] \ket{1} \nonumber \\
\hat{D}[\alpha(t)] &= \exp [\alpha(t)\hat{a}^\dagger-\alpha(t)^*\hat{a}],
\end{align}

\noindent where $\ket{n}$ designates the $n$-th number state of a harmonic oscillator with frequency $\omega_y$ as defined above, and $\hat{a},\hat{a}^\dagger$ the corresponding ladder operators.
The complex displacement $\alpha(t)$ corresponds to the origin $\tilde{y}(t),\tilde{k}_y(t)$ of the co-oscillating frame in phase space, with
\begin{align*}
\tilde{y} = \sqrt{2} r_{0,y} \Re [\alpha(t)], \qquad
\tilde{k}_y = \sqrt{2} r_{0,y}^{-1} \Im [\alpha(t)].
\end{align*}

The question remains how to appropriately define the co-oscillating frame $\tilde{y}(t),\tilde{k}_y(t)$ in our analysis.
A stringent means to obtain its origin is readily provided by observing the average transversal momentum of the emitted atoms, which are in state $\ket{g,\pm p}$ that by construction has the same transversal center-of-mass as $\ket{g,0}$ (see fig.~\ref{fig:image}, inset), hence fully defining the collective oscillation.
However, during the first phase of excitation (especially for weak excitation scalings) when the emitted atom number is low, the uncertainty of the experimentally determined center-of-mass is large.
Moreover, at later times the emission increasingly perturbs the transversal dynamics, making a direct extraction of the oscillating frame from experimental data inconsistent with our model, which omits emission.

\begin{figure}
	\includegraphics{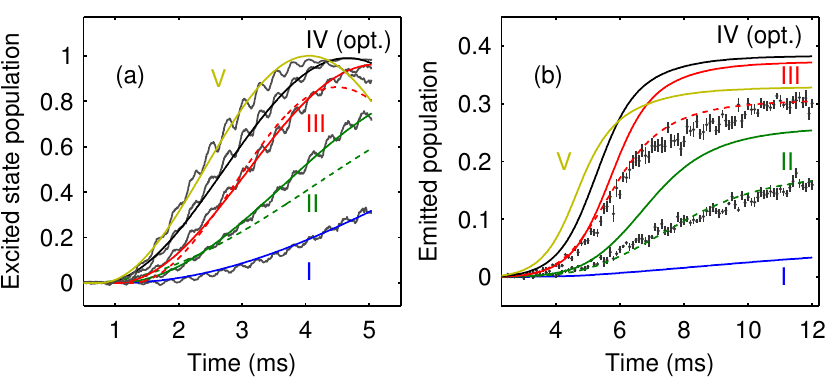}
	\caption{(Color online) (a) Grey lines: populations of the excited state $\eta$ extracted from the GPE result within the two-state superposition model. Solid lines: fits of a constant Rabi coupling model to the numerical result. Dashed lines: populations for Rabi coupling parameters that were used as input to the emission calculations for intermediate excitation strength (green, red). (b) Resulting emission dynamics curves, corresponding to the excitation curves shown in (a), respectively. Experimental data for the intermediate scalings (II and III) are shown as dots.}
	\label{fig:exc}
\end{figure}

Instead, we proceed by numerically optimizing the phase space trajectory $\alpha(t)$ of the co-oscillating frame origin, at each time $t$ maximizing the overlap $\chi[\alpha(t)]$ of the GPE result~$\ket{\psi(t)}$ with a superposition of the basis states:

\begin{align*}
\chi[\alpha(t)] = \left| \bra{\psi(t)} \hat{D}[\alpha(t)] \left\lbrace \sqrt{1-\eta(t)} \ket{0} + \sqrt{\eta(t)} \ket{1} \right\rbrace \right|^2,
\end{align*}
with the excited state population $\eta(t)$ as free parameter.
For optimal excitation, this procedure yields $\chi \gtrsim 0.95$ overlap at all times $t$.
In fig.~\ref{fig:transdyn}d, a comparison of the numerically optimized $\tilde{k}_y$ and the center-of-mass position of the emitted atoms is shown. 
The excellent agreement at times $t\lesssim\SI{5}{ms}$ strongly corroborates the displaced oscillator eigenstate model.
In fig.~\ref{fig:transdyn}c, the momentum distribution of the two-state superposition is shown, reflecting the smooth transition from ground to excited state, with strong beating between the states at intermediate excited fractions. 
Fig.~\ref{fig:exc}(a) (grey lines) shows the population of the excited state $\eta(t)$, for the five excitation strengths inferred from comparison of the GPE result to the experimental data~(fig.~\ref{fig:transdyn}).
The smooth transitions observed in the co-oscillating frame are in stark contrast to the strong transient population of higher states (as e.g. shown in~\cite{Jirari2009,Khani2009}) which would be obtained by simple projection of the GPE result on the oscillator eigenfunctions.

To facilitate the inclusion of continuous pumping into the emission dynamics calculation, we approximate the excitation process by a constant, near-resonant Rabi coupling with coupling strength $\Omega$, detuning $\delta$ and initial time delay $t_0$, and obtain
\begin{align}
\hat{H}_\mathrm{pump} (t) &= \left\{ 
\begin{array}{ll}
	\frac{\hbar\tilde{\Omega}}{2}   \ket{e,0}\bra{g,0} + \textrm{H.c.} &: t_0 \leq t \leq \SI{5}{ms} \\
	0 &: \mathrm{otherwise}
\end{array}
\right. \nonumber \\
\tilde{\Omega} &= \Omega \cdot \exp \left[ -i\left(E_{y,1}/\hbar + \delta\right) (t-t_0) \right]
\end{align}
for the pumping Hamiltonian in eq.~\eqref{eq:twinproduction}.
This corresponds to an approximation $\tilde{\eta}(t)$ to the numerically obtained two-state populations:
\begin{equation}
\tilde{\eta} (t) = \left\{ 
\begin{array}{ll}
	\left(\frac{\Omega}{\Omega'}\right)^2 \sin^2 \left[\Omega' (t-t_0) \right] &: t_0 \leq t \leq \SI{5}{ms} \\
	0 &: \mathrm{otherwise}
\end{array}
\right.
\label{eq:rabi}
\end{equation}
where $\Omega' = \sqrt{\Omega^2 + \delta^2}$. 
Fig.~\ref{fig:exc}(a) (solid lines) shows, that eq.~\eqref{eq:rabi} is an appropriate approximation to the numerically obtained $\eta(t)$, if $\Omega, \Delta$ and $t_0$ are left as fit parameters.

For the optimal, weakest and strongest excitations (lines IV, I, and V in fig.~\ref{fig:exc}, respectively), using the fitted parameters as input to the dynamics calculation directly yields excellent agreement of the emitted atom number to the experiment as shown in fig.~\ref{fig:pop}.
Only for the two intermediate strengths (lines II and III), emission curves derived in this way show unsatisfactory agreement (solid lines II and III in fig.~\ref{fig:exc}(b) ).
We attribute this to failure of the two-level approximation to yield estimates for the $\eta(t)$ that are sufficiently precise for prediction of the Bose-enhanced emission dynamics, which is highly sensitive to changes in the initial source population.
To be able to verify the validity of our emission dynamics theory for intermediate excitations, only for those two settings we slightly modified the parameters used as input to the emission calculation.
The resulting source populations $\tilde{\eta}(t)$ are shown as dashed lines in fig.~\ref{fig:exc}(a).
Even the slight deviation from the numerically obtained $\eta(t)$ is sufficient to considerably change the emission dynamics results and reach agreement to the experiment (dashed lines in fig.~\ref{fig:exc}(b) ).

A more detailed description of the transversal dynamics would additionally require the explicit treatment of the orbital dynamics, e.g. within the multi-configurational Hartree method for bosons \cite{alon:08,grond.pra:09b}, which is beyond the scope of our present paper.

\section{Calculation of relative variances}
\label{sec:variance}

In fig.~\ref{fig:pop} (inset), we show the fluctuations of the relative population of the twin-beam states in terms of the relative variance $\xi^2$.
It can be shown that the fluctuations of the emitted number of atoms $N$ can be written with respect to the variance of a Poissonian distribution as

\begin{equation*}
\textrm{Var}(N-pn) \approx 2 \xi^2 \bar{N}.
\end{equation*}
In this and the following equations, $\bar{X}$ denotes the average of a random variable $X$.
We eliminated the contribution of fluctuations of the total atom number $n$ by subtracting the conditional expected value of $N$ for a given total atom number $\textrm{E}(N|n)=pn$, where $p$ denotes the emission probability.
The latter can be estimated from the mean emitted fraction of atoms $p\approx\bar{N}/\bar{n}$.
The factor of 2 arises due to the atoms always being emitted pairwise.
Note, that assuming a Poissonian distribution for the shot noise of uncorrelated spontaneous emission is generally only valid within the approximation of negligible source depletion ($N \ll n_0$, $n_0$ denoting the population of the source state), i.e., at early times during the emission.
Also, we have to neglect fluctuations of the relative source population $n_0/n$.
Those would result from fluctuations in temperature and excitation efficiency, which are inaccessible to independent characterization.
Hence, an interpretation of the relative variance $\xi^2$ as being normalized to the shot noise of randomly emitted atoms for an ensemble of identical initial conditions is rather crude.
Still it gives the correct order of magnitude, and allows for comparison between different times during emission, and different excitation strengths.

In the next step, the contribution of photon shot noise in the fluorescence imaging~\cite{Buecker2009} has to be accounted for.
This can be done in a similar manner to the calculation shown in~\cite{buecker:11} (Supplementary Information), and leads to:

\begin{equation*}
\textrm{Var}(S-ps) \approx 2 \xi^2 m \bar{S} + 2\bar{S} + \sigma_b^2.
\end{equation*}
In this equation, $s$ and $S$ denote the total number of photons scattered from atoms, and the number of photons outside the source region as defined above, respectively.
The average number of detected photos emitted by each atom is denoted as $m$, typically $m\approx 12$.
This means, that $\bar{S}=m\bar{N},\bar{s}=m\bar{n}$, and thus $\bar{S}/\bar{s}=\bar{N}/\bar{n}\approx p$. 
However, $\textrm{Var}(S-ps) > \textrm{Var}(mN-pmn)$.
The second term represents the photon shot noise of light scattered by the atoms, where the factor of 2 arises due to the EMCCD camera used for imaging~\cite{Buecker2009}.
Background light is taken into account by $\sigma_b^2$, denoting the variance of the number of background photons originating from stray light and detector noise within the twin-beam regions.


\begin{thebibliography}{52}%
\makeatletter
\providecommand \@ifxundefined [1]{%
 \@ifx{#1\undefined}
}%
\providecommand \@ifnum [1]{%
 \ifnum #1\expandafter \@firstoftwo
 \else \expandafter \@secondoftwo
 \fi
}%
\providecommand \@ifx [1]{%
 \ifx #1\expandafter \@firstoftwo
 \else \expandafter \@secondoftwo
 \fi
}%
\providecommand \natexlab [1]{#1}%
\providecommand \enquote  [1]{``#1''}%
\providecommand \bibnamefont  [1]{#1}%
\providecommand \bibfnamefont [1]{#1}%
\providecommand \citenamefont [1]{#1}%
\providecommand \href@noop [0]{\@secondoftwo}%
\providecommand \href [0]{\begingroup \@sanitize@url \@href}%
\providecommand \@href[1]{\@@startlink{#1}\@@href}%
\providecommand \@@href[1]{\endgroup#1\@@endlink}%
\providecommand \@sanitize@url [0]{\catcode `\\12\catcode `\$12\catcode
  `\&12\catcode `\#12\catcode `\^12\catcode `\_12\catcode `\%12\relax}%
\providecommand \@@startlink[1]{}%
\providecommand \@@endlink[0]{}%
\providecommand \url  [0]{\begingroup\@sanitize@url \@url }%
\providecommand \@url [1]{\endgroup\@href {#1}{\urlprefix }}%
\providecommand \urlprefix  [0]{URL }%
\providecommand \Eprint [0]{\href }%
\@ifxundefined \urlstyle {%
  \providecommand \doi  [0]{\begingroup \@sanitize@url \@doi}%
  \providecommand \@doi [1]{\endgroup \@@startlink {\doibase
  #1}doi:\discretionary {}{}{}#1\@@endlink }%
}{%
  \providecommand \doi  [0]{doi:\discretionary{}{}{}\begingroup
  \urlstyle{rm}\Url }%
}%
\providecommand \doibase [0]{http://dx.doi.org/}%
\providecommand \Doi [0]{\begingroup \@sanitize@url \@Doi }%
\providecommand \@Doi  [1]{\endgroup\@@startlink{\doibase#1}\@@Doi}%
\providecommand \@@Doi [1]{#1\@@endlink}%
\providecommand \selectlanguage [0]{\@gobble}%
\providecommand \bibinfo  [0]{\@secondoftwo}%
\providecommand \bibfield  [0]{\@secondoftwo}%
\providecommand \translation [1]{[#1]}%
\providecommand \BibitemOpen [0]{}%
\providecommand \bibitemStop [0]{}%
\providecommand \bibitemNoStop [0]{.\EOS\space}%
\providecommand \EOS [0]{\spacefactor3000\relax}%
\providecommand \BibitemShut  [1]{\csname bibitem#1\endcsname}%
\bibitem [{\citenamefont {Burnham}\ and\ \citenamefont
  {Weinberg}(1970)}]{Burnham1970}%
  \BibitemOpen
  \bibfield  {author} {\bibinfo {author} {\bibfnamefont {D.}~\bibnamefont
  {Burnham}}\ and\ \bibinfo {author} {\bibfnamefont {D.}~\bibnamefont
  {Weinberg}},\ }\href@noop {} {\bibfield  {journal} {\bibinfo  {journal}
  {Physical Review Letters},\ }\textbf {\bibinfo {volume} {25}},\ \bibinfo
  {pages} {84} (\bibinfo {year} {1970})}\BibitemShut {NoStop}%
\bibitem [{\citenamefont {Heidmann}\ \emph {et~al.}(1987)\citenamefont
  {Heidmann}, \citenamefont {Horowicz}, \citenamefont {Reynaud}, \citenamefont
  {Giacobino}, \citenamefont {Fabre},\ and\ \citenamefont
  {Camy}}]{Heidmann1987}%
  \BibitemOpen
  \bibfield  {author} {\bibinfo {author} {\bibfnamefont {A.}~\bibnamefont
  {Heidmann}}, \bibinfo {author} {\bibfnamefont {R.}~\bibnamefont {Horowicz}},
  \bibinfo {author} {\bibfnamefont {S.}~\bibnamefont {Reynaud}}, \bibinfo
  {author} {\bibfnamefont {E.}~\bibnamefont {Giacobino}}, \bibinfo {author}
  {\bibfnamefont {C.}~\bibnamefont {Fabre}}, \ and\ \bibinfo {author}
  {\bibfnamefont {G.}~\bibnamefont {Camy}},\ }\href@noop {} {\bibfield
  {journal} {\bibinfo  {journal} {Physical Review Letters},\ }\textbf {\bibinfo
  {volume} {59}},\ \bibinfo {pages} {2555} (\bibinfo {year}
  {1987})}\BibitemShut {NoStop}%
\bibitem [{\citenamefont {Chikkatur}\ \emph {et~al.}(2000)\citenamefont
  {Chikkatur}, \citenamefont {G\"{o}rlitz}, \citenamefont {Stamper-Kurn},
  \citenamefont {Inouye}, \citenamefont {Gupta},\ and\ \citenamefont
  {Ketterle}}]{Chikkatur2000}%
  \BibitemOpen
  \bibfield  {author} {\bibinfo {author} {\bibfnamefont {A.}~\bibnamefont
  {Chikkatur}}, \bibinfo {author} {\bibfnamefont {A.}~\bibnamefont
  {G\"{o}rlitz}}, \bibinfo {author} {\bibfnamefont {D.}~\bibnamefont
  {Stamper-Kurn}}, \bibinfo {author} {\bibfnamefont {S.}~\bibnamefont
  {Inouye}}, \bibinfo {author} {\bibfnamefont {S.}~\bibnamefont {Gupta}}, \
  and\ \bibinfo {author} {\bibfnamefont {W.}~\bibnamefont {Ketterle}},\
  }\href@noop {} {\bibfield  {journal} {\bibinfo  {journal} {Physical Review
  Letters},\ }\textbf {\bibinfo {volume} {85}},\ \bibinfo {pages} {483 }
  (\bibinfo {year} {2000})}\BibitemShut {NoStop}%
\bibitem [{\citenamefont {Perrin}\ \emph {et~al.}(2007)\citenamefont {Perrin},
  \citenamefont {Chang}, \citenamefont {Krachmalnicoff}, \citenamefont
  {Schellekens}, \citenamefont {Boiron}, \citenamefont {Aspect},\ and\
  \citenamefont {Westbrook}}]{Perrin2007}%
  \BibitemOpen
  \bibfield  {author} {\bibinfo {author} {\bibfnamefont {A.}~\bibnamefont
  {Perrin}}, \bibinfo {author} {\bibfnamefont {H.}~\bibnamefont {Chang}},
  \bibinfo {author} {\bibfnamefont {V.}~\bibnamefont {Krachmalnicoff}},
  \bibinfo {author} {\bibfnamefont {M.}~\bibnamefont {Schellekens}}, \bibinfo
  {author} {\bibfnamefont {D.}~\bibnamefont {Boiron}}, \bibinfo {author}
  {\bibfnamefont {A.}~\bibnamefont {Aspect}}, \ and\ \bibinfo {author}
  {\bibfnamefont {C.}~\bibnamefont {Westbrook}},\ }\href@noop {} {\bibfield
  {journal} {\bibinfo  {journal} {Physical Review Letters},\ }\textbf {\bibinfo
  {volume} {99}},\ \bibinfo {pages} {150405} (\bibinfo {year}
  {2007})}\BibitemShut {NoStop}%
\bibitem [{\citenamefont {Jaskula}\ \emph {et~al.}(2010)\citenamefont
  {Jaskula}, \citenamefont {Bonneau}, \citenamefont {Partridge}, \citenamefont
  {Krachmalnicoff}, \citenamefont {Deuar}, \citenamefont {Kheruntsyan},
  \citenamefont {Aspect}, \citenamefont {Boiron},\ and\ \citenamefont
  {Westbrook}}]{Jaskula2010a}%
  \BibitemOpen
  \bibfield  {author} {\bibinfo {author} {\bibfnamefont {J.-C.}\ \bibnamefont
  {Jaskula}}, \bibinfo {author} {\bibfnamefont {M.}~\bibnamefont {Bonneau}},
  \bibinfo {author} {\bibfnamefont {G.}~\bibnamefont {Partridge}}, \bibinfo
  {author} {\bibfnamefont {V.}~\bibnamefont {Krachmalnicoff}}, \bibinfo
  {author} {\bibfnamefont {P.}~\bibnamefont {Deuar}}, \bibinfo {author}
  {\bibfnamefont {K.}~\bibnamefont {Kheruntsyan}}, \bibinfo {author}
  {\bibfnamefont {A.}~\bibnamefont {Aspect}}, \bibinfo {author} {\bibfnamefont
  {D.}~\bibnamefont {Boiron}}, \ and\ \bibinfo {author} {\bibfnamefont
  {C.}~\bibnamefont {Westbrook}},\ }\href@noop {} {\bibfield  {journal}
  {\bibinfo  {journal} {Physical Review Letters},\ }\textbf {\bibinfo {volume}
  {105}},\ \bibinfo {pages} {190402} (\bibinfo {year} {2010})}\BibitemShut
  {NoStop}%
\bibitem [{\citenamefont {Deng}\ \emph {et~al.}(1999)\citenamefont {Deng},
  \citenamefont {Hagley}, \citenamefont {Wen}, \citenamefont {Trippenbach},
  \citenamefont {Band}, \citenamefont {Julienne}, \citenamefont {Simsarian},
  \citenamefont {Helmerson}, \citenamefont {Rolston},\ and\ \citenamefont
  {Phillips}}]{Deng1999}%
  \BibitemOpen
  \bibfield  {author} {\bibinfo {author} {\bibfnamefont {L.}~\bibnamefont
  {Deng}}, \bibinfo {author} {\bibfnamefont {E.~W.}\ \bibnamefont {Hagley}},
  \bibinfo {author} {\bibfnamefont {J.}~\bibnamefont {Wen}}, \bibinfo {author}
  {\bibfnamefont {M.}~\bibnamefont {Trippenbach}}, \bibinfo {author}
  {\bibfnamefont {Y.}~\bibnamefont {Band}}, \bibinfo {author} {\bibfnamefont
  {P.~S.}\ \bibnamefont {Julienne}}, \bibinfo {author} {\bibfnamefont {J.~E.}\
  \bibnamefont {Simsarian}}, \bibinfo {author} {\bibfnamefont {K.}~\bibnamefont
  {Helmerson}}, \bibinfo {author} {\bibfnamefont {S.~L.}\ \bibnamefont
  {Rolston}}, \ and\ \bibinfo {author} {\bibfnamefont {W.~D.}\ \bibnamefont
  {Phillips}},\ }\href@noop {} {\bibfield  {journal} {\bibinfo  {journal}
  {Nature},\ }\textbf {\bibinfo {volume} {398}},\ \bibinfo {pages} {218}
  (\bibinfo {year} {1999})}\BibitemShut {NoStop}%
\bibitem [{\citenamefont {Vogels}\ \emph {et~al.}(2002)\citenamefont {Vogels},
  \citenamefont {Xu},\ and\ \citenamefont {Ketterle}}]{Vogels2002}%
  \BibitemOpen
  \bibfield  {author} {\bibinfo {author} {\bibfnamefont {J.}~\bibnamefont
  {Vogels}}, \bibinfo {author} {\bibfnamefont {K.}~\bibnamefont {Xu}}, \ and\
  \bibinfo {author} {\bibfnamefont {W.}~\bibnamefont {Ketterle}},\ }\href@noop
  {} {\bibfield  {journal} {\bibinfo  {journal} {Physical Review Letters},\
  }\textbf {\bibinfo {volume} {89}},\ \bibinfo {pages} {020401} (\bibinfo
  {year} {2002})}\BibitemShut {NoStop}%
\bibitem [{\citenamefont {Gemelke}\ \emph {et~al.}(2005)\citenamefont
  {Gemelke}, \citenamefont {Sarajlic}, \citenamefont {Bidel}, \citenamefont
  {Hong},\ and\ \citenamefont {Chu}}]{Gemelke2005}%
  \BibitemOpen
  \bibfield  {author} {\bibinfo {author} {\bibfnamefont {N.}~\bibnamefont
  {Gemelke}}, \bibinfo {author} {\bibfnamefont {E.}~\bibnamefont {Sarajlic}},
  \bibinfo {author} {\bibfnamefont {Y.}~\bibnamefont {Bidel}}, \bibinfo
  {author} {\bibfnamefont {S.}~\bibnamefont {Hong}}, \ and\ \bibinfo {author}
  {\bibfnamefont {S.}~\bibnamefont {Chu}},\ }\href@noop {} {\bibfield
  {journal} {\bibinfo  {journal} {Physical Review Letters},\ }\textbf {\bibinfo
  {volume} {95}},\ \bibinfo {pages} {170404} (\bibinfo {year}
  {2005})}\BibitemShut {NoStop}%
\bibitem [{\citenamefont {Campbell}\ \emph {et~al.}(2006)\citenamefont
  {Campbell}, \citenamefont {Mun}, \citenamefont {Boyd}, \citenamefont
  {Streed}, \citenamefont {Ketterle},\ and\ \citenamefont
  {Pritchard}}]{Campbell2006}%
  \BibitemOpen
  \bibfield  {author} {\bibinfo {author} {\bibfnamefont {G.~K.}\ \bibnamefont
  {Campbell}}, \bibinfo {author} {\bibfnamefont {J.}~\bibnamefont {Mun}},
  \bibinfo {author} {\bibfnamefont {M.}~\bibnamefont {Boyd}}, \bibinfo {author}
  {\bibfnamefont {E.~W.}\ \bibnamefont {Streed}}, \bibinfo {author}
  {\bibfnamefont {W.}~\bibnamefont {Ketterle}}, \ and\ \bibinfo {author}
  {\bibfnamefont {D.~E.}\ \bibnamefont {Pritchard}},\ }\href@noop {} {\bibfield
   {journal} {\bibinfo  {journal} {Physical Review Letters},\ }\textbf
  {\bibinfo {volume} {96}},\ \bibinfo {pages} {020406} (\bibinfo {year}
  {2006})}\BibitemShut {NoStop}%
\bibitem [{\citenamefont {Dall}\ \emph {et~al.}(2009)\citenamefont {Dall},
  \citenamefont {Byron}, \citenamefont {Truscott}, \citenamefont {Dennis},
  \citenamefont {Johnsson},\ and\ \citenamefont {Hope}}]{Dall2009}%
  \BibitemOpen
  \bibfield  {author} {\bibinfo {author} {\bibfnamefont {R.~G.}\ \bibnamefont
  {Dall}}, \bibinfo {author} {\bibfnamefont {L.~J.}\ \bibnamefont {Byron}},
  \bibinfo {author} {\bibfnamefont {A.~G.}\ \bibnamefont {Truscott}}, \bibinfo
  {author} {\bibfnamefont {G.~R.}\ \bibnamefont {Dennis}}, \bibinfo {author}
  {\bibfnamefont {M.~T.}\ \bibnamefont {Johnsson}}, \ and\ \bibinfo {author}
  {\bibfnamefont {J.~J.}\ \bibnamefont {Hope}},\ }\href@noop {} {\bibfield
  {journal} {\bibinfo  {journal} {Physical Review A},\ }\textbf {\bibinfo
  {volume} {79}},\ \bibinfo {pages} {011601} (\bibinfo {year}
  {2009})}\BibitemShut {NoStop}%
\bibitem [{\citenamefont {RuGway}\ \emph {et~al.}(2011)\citenamefont {RuGway},
  \citenamefont {Hodgman}, \citenamefont {Dall}, \citenamefont {Johnson},\ and\
  \citenamefont {Truscott}}]{RuGway2011}%
  \BibitemOpen
  \bibfield  {author} {\bibinfo {author} {\bibfnamefont {W.}~\bibnamefont
  {RuGway}}, \bibinfo {author} {\bibfnamefont {S.}~\bibnamefont {Hodgman}},
  \bibinfo {author} {\bibfnamefont {R.}~\bibnamefont {Dall}}, \bibinfo {author}
  {\bibfnamefont {M.}~\bibnamefont {Johnson}}, \ and\ \bibinfo {author}
  {\bibfnamefont {A.}~\bibnamefont {Truscott}},\ }\href@noop {} {\bibfield
  {journal} {\bibinfo  {journal} {Physical Review Letters},\ }\textbf {\bibinfo
  {volume} {107}},\ \bibinfo {pages} {075301} (\bibinfo {year}
  {2011})}\BibitemShut {NoStop}%
\bibitem [{\citenamefont {B{\"u}cker}\ \emph {et~al.}(2011)\citenamefont
  {B{\"u}cker}, \citenamefont {Grond}, \citenamefont {Manz}, \citenamefont
  {Berrada}, \citenamefont {Betz}, \citenamefont {Koller}, \citenamefont
  {Hohenester}, \citenamefont {Schumm}, \citenamefont {Perrin},\ and\
  \citenamefont {Schmiedmayer}}]{buecker:11}%
  \BibitemOpen
  \bibfield  {author} {\bibinfo {author} {\bibfnamefont {R.}~\bibnamefont
  {B{\"u}cker}}, \bibinfo {author} {\bibfnamefont {J.}~\bibnamefont {Grond}},
  \bibinfo {author} {\bibfnamefont {S.}~\bibnamefont {Manz}}, \bibinfo {author}
  {\bibfnamefont {T.}~\bibnamefont {Berrada}}, \bibinfo {author} {\bibfnamefont
  {T.}~\bibnamefont {Betz}}, \bibinfo {author} {\bibfnamefont {C.}~\bibnamefont
  {Koller}}, \bibinfo {author} {\bibfnamefont {U.}~\bibnamefont {Hohenester}},
  \bibinfo {author} {\bibfnamefont {T.}~\bibnamefont {Schumm}}, \bibinfo
  {author} {\bibfnamefont {A.}~\bibnamefont {Perrin}}, \ and\ \bibinfo {author}
  {\bibfnamefont {J.}~\bibnamefont {Schmiedmayer}},\ }\href@noop {} {\bibfield
  {journal} {\bibinfo  {journal} {Nature Physics},\ }\textbf {\bibinfo {volume}
  {7}},\ \bibinfo {pages} {508} (\bibinfo {year} {2011})}\BibitemShut {NoStop}%
\bibitem [{\citenamefont {Klempt}\ \emph {et~al.}(2010)\citenamefont {Klempt},
  \citenamefont {Topic}, \citenamefont {Gebreyesus}, \citenamefont {Scherer},
  \citenamefont {Henninger}, \citenamefont {Hyllus}, \citenamefont {Ertmer},
  \citenamefont {Santos},\ and\ \citenamefont {Arlt}}]{Klempt2010}%
  \BibitemOpen
  \bibfield  {author} {\bibinfo {author} {\bibfnamefont {C.}~\bibnamefont
  {Klempt}}, \bibinfo {author} {\bibfnamefont {O.}~\bibnamefont {Topic}},
  \bibinfo {author} {\bibfnamefont {G.}~\bibnamefont {Gebreyesus}}, \bibinfo
  {author} {\bibfnamefont {M.}~\bibnamefont {Scherer}}, \bibinfo {author}
  {\bibfnamefont {T.}~\bibnamefont {Henninger}}, \bibinfo {author}
  {\bibfnamefont {P.}~\bibnamefont {Hyllus}}, \bibinfo {author} {\bibfnamefont
  {W.}~\bibnamefont {Ertmer}}, \bibinfo {author} {\bibfnamefont
  {L.}~\bibnamefont {Santos}}, \ and\ \bibinfo {author} {\bibfnamefont {J.~J.}\
  \bibnamefont {Arlt}},\ }\href@noop {} {\bibfield  {journal} {\bibinfo
  {journal} {Physical Review Letters},\ }\textbf {\bibinfo {volume} {104}},\
  \bibinfo {pages} {195303} (\bibinfo {year} {2010})}\BibitemShut {NoStop}%
\bibitem [{\citenamefont {L\"{u}cke}\ \emph {et~al.}(2011)\citenamefont
  {L\"{u}cke}, \citenamefont {Scherer}, \citenamefont {Kruse}, \citenamefont
  {Pezz\'{e}}, \citenamefont {Deuretzbacher}, \citenamefont {Hyllus},
  \citenamefont {Topic}, \citenamefont {Peise}, \citenamefont {Ertmer},
  \citenamefont {Arlt}, \citenamefont {Santos}, \citenamefont {Smerzi},\ and\
  \citenamefont {Klempt}}]{Lucke2011}%
  \BibitemOpen
  \bibfield  {author} {\bibinfo {author} {\bibfnamefont {B.}~\bibnamefont
  {L\"{u}cke}}, \bibinfo {author} {\bibfnamefont {M.}~\bibnamefont {Scherer}},
  \bibinfo {author} {\bibfnamefont {J.}~\bibnamefont {Kruse}}, \bibinfo
  {author} {\bibfnamefont {L.}~\bibnamefont {Pezz\'{e}}}, \bibinfo {author}
  {\bibfnamefont {F.}~\bibnamefont {Deuretzbacher}}, \bibinfo {author}
  {\bibfnamefont {P.}~\bibnamefont {Hyllus}}, \bibinfo {author} {\bibfnamefont
  {O.}~\bibnamefont {Topic}}, \bibinfo {author} {\bibfnamefont
  {J.}~\bibnamefont {Peise}}, \bibinfo {author} {\bibfnamefont
  {W.}~\bibnamefont {Ertmer}}, \bibinfo {author} {\bibfnamefont
  {J.}~\bibnamefont {Arlt}}, \bibinfo {author} {\bibfnamefont {L.}~\bibnamefont
  {Santos}}, \bibinfo {author} {\bibfnamefont {A.}~\bibnamefont {Smerzi}}, \
  and\ \bibinfo {author} {\bibfnamefont {C.}~\bibnamefont {Klempt}},\
  }\href@noop {} {\bibfield  {journal} {\bibinfo  {journal} {Science},\
  }\textbf {\bibinfo {volume} {334}},\ \bibinfo {pages} {773} (\bibinfo {year}
  {2011})}\BibitemShut {NoStop}%
\bibitem [{\citenamefont {Bookjans}\ \emph {et~al.}(2011)\citenamefont
  {Bookjans}, \citenamefont {Hamley},\ and\ \citenamefont
  {Chapman}}]{Bookjans2011}%
  \BibitemOpen
  \bibfield  {author} {\bibinfo {author} {\bibfnamefont {E.~M.}\ \bibnamefont
  {Bookjans}}, \bibinfo {author} {\bibfnamefont {C.~D.}\ \bibnamefont
  {Hamley}}, \ and\ \bibinfo {author} {\bibfnamefont {M.~S.}\ \bibnamefont
  {Chapman}},\ }\href@noop {} {\bibfield  {journal} {\bibinfo  {journal}
  {Physical Review Letters},\ }\textbf {\bibinfo {volume} {107}},\ \bibinfo
  {pages} {210406} (\bibinfo {year} {2011})}\BibitemShut {NoStop}%
\bibitem [{\citenamefont {Hamley}\ \emph {et~al.}(2012)\citenamefont {Hamley},
  \citenamefont {Gerving}, \citenamefont {Hoang}, \citenamefont {Bookjans},\
  and\ \citenamefont {Chapman}}]{Hamley2012}%
  \BibitemOpen
  \bibfield  {author} {\bibinfo {author} {\bibfnamefont {C.~D.}\ \bibnamefont
  {Hamley}}, \bibinfo {author} {\bibfnamefont {C.~S.}\ \bibnamefont {Gerving}},
  \bibinfo {author} {\bibfnamefont {T.~M.}\ \bibnamefont {Hoang}}, \bibinfo
  {author} {\bibfnamefont {E.~M.}\ \bibnamefont {Bookjans}}, \ and\ \bibinfo
  {author} {\bibfnamefont {M.~S.}\ \bibnamefont {Chapman}},\ }\href@noop {}
  {\bibfield  {journal} {\bibinfo  {journal} {Nature Physics},\ }\textbf
  {\bibinfo {volume} {8}},\ \bibinfo {pages} {305} (\bibinfo {year}
  {2012})}\BibitemShut {NoStop}%
\bibitem [{\citenamefont {Gross}\ \emph {et~al.}(2011)\citenamefont {Gross},
  \citenamefont {Strobel}, \citenamefont {Nicklas}, \citenamefont {Zibold},
  \citenamefont {Bar-Gill}, \citenamefont {Kurizki},\ and\ \citenamefont
  {Oberthaler}}]{Gross2011}%
  \BibitemOpen
  \bibfield  {author} {\bibinfo {author} {\bibfnamefont {C.}~\bibnamefont
  {Gross}}, \bibinfo {author} {\bibfnamefont {H.}~\bibnamefont {Strobel}},
  \bibinfo {author} {\bibfnamefont {E.}~\bibnamefont {Nicklas}}, \bibinfo
  {author} {\bibfnamefont {T.}~\bibnamefont {Zibold}}, \bibinfo {author}
  {\bibfnamefont {N.}~\bibnamefont {Bar-Gill}}, \bibinfo {author}
  {\bibfnamefont {G.}~\bibnamefont {Kurizki}}, \ and\ \bibinfo {author}
  {\bibfnamefont {M.~K.}\ \bibnamefont {Oberthaler}},\ }\href@noop {}
  {\bibfield  {journal} {\bibinfo  {journal} {Nature},\ }\textbf {\bibinfo
  {volume} {480}},\ \bibinfo {pages} {290} (\bibinfo {year}
  {2011})}\BibitemShut {NoStop}%
\bibitem [{\citenamefont {Perrin}\ \emph {et~al.}(2012)\citenamefont {Perrin},
  \citenamefont {B\"{u}cker}, \citenamefont {Manz}, \citenamefont {Betz},
  \citenamefont {Koller}, \citenamefont {Plisson}, \citenamefont {Schumm},\
  and\ \citenamefont {Schmiedmayer}}]{Perrin2012}%
  \BibitemOpen
  \bibfield  {author} {\bibinfo {author} {\bibfnamefont {A.}~\bibnamefont
  {Perrin}}, \bibinfo {author} {\bibfnamefont {R.}~\bibnamefont {B\"{u}cker}},
  \bibinfo {author} {\bibfnamefont {S.}~\bibnamefont {Manz}}, \bibinfo {author}
  {\bibfnamefont {T.}~\bibnamefont {Betz}}, \bibinfo {author} {\bibfnamefont
  {C.}~\bibnamefont {Koller}}, \bibinfo {author} {\bibfnamefont
  {T.}~\bibnamefont {Plisson}}, \bibinfo {author} {\bibfnamefont
  {T.}~\bibnamefont {Schumm}}, \ and\ \bibinfo {author} {\bibfnamefont
  {J.}~\bibnamefont {Schmiedmayer}},\ }\href@noop {} {\bibfield  {journal}
  {\bibinfo  {journal} {Nature Physics},\ }\textbf {\bibinfo {volume} {8}},\
  \bibinfo {pages} {195} (\bibinfo {year} {2012})}\BibitemShut {NoStop}%
\bibitem [{\citenamefont {Vardi}\ and\ \citenamefont
  {Anglin}(2001)}]{vardi:01}%
  \BibitemOpen
  \bibfield  {author} {\bibinfo {author} {\bibfnamefont {A.}~\bibnamefont
  {Vardi}}\ and\ \bibinfo {author} {\bibfnamefont {J.~R.}\ \bibnamefont
  {Anglin}},\ }\href@noop {} {\bibfield  {journal} {\bibinfo  {journal}
  {Physical Review Letters},\ }\textbf {\bibinfo {volume} {86}},\ \bibinfo
  {pages} {568} (\bibinfo {year} {2001})}\BibitemShut {NoStop}%
\bibitem [{\citenamefont {Pu}\ and\ \citenamefont {Meystre}(2000)}]{Pu2000}%
  \BibitemOpen
  \bibfield  {author} {\bibinfo {author} {\bibfnamefont {H.}~\bibnamefont
  {Pu}}\ and\ \bibinfo {author} {\bibfnamefont {P.}~\bibnamefont {Meystre}},\
  }\href@noop {} {\bibfield  {journal} {\bibinfo  {journal} {Physical Review
  Letters},\ }\textbf {\bibinfo {volume} {85}},\ \bibinfo {pages} {3987}
  (\bibinfo {year} {2000})}\BibitemShut {NoStop}%
\bibitem [{\citenamefont {Duan}\ \emph {et~al.}(2000)\citenamefont {Duan},
  \citenamefont {Sorensen}, \citenamefont {Cirac},\ and\ \citenamefont
  {Zoller}}]{Duan2000a}%
  \BibitemOpen
  \bibfield  {author} {\bibinfo {author} {\bibfnamefont {L.-M.}\ \bibnamefont
  {Duan}}, \bibinfo {author} {\bibfnamefont {A.}~\bibnamefont {Sorensen}},
  \bibinfo {author} {\bibfnamefont {J.~I.}\ \bibnamefont {Cirac}}, \ and\
  \bibinfo {author} {\bibfnamefont {P.}~\bibnamefont {Zoller}},\ }\href@noop {}
  {\bibfield  {journal} {\bibinfo  {journal} {Physical Review Letters},\
  }\textbf {\bibinfo {volume} {85}},\ \bibinfo {pages} {3991} (\bibinfo {year}
  {2000})}\BibitemShut {NoStop}%
\bibitem [{\citenamefont {Bach}\ \emph {et~al.}(2002)\citenamefont {Bach},
  \citenamefont {Trippenbach},\ and\ \citenamefont
  {Rz\k{a}\.{z}ewski}}]{Bach2002}%
  \BibitemOpen
  \bibfield  {author} {\bibinfo {author} {\bibfnamefont {R.}~\bibnamefont
  {Bach}}, \bibinfo {author} {\bibfnamefont {M.}~\bibnamefont {Trippenbach}}, \
  and\ \bibinfo {author} {\bibfnamefont {K.}~\bibnamefont
  {Rz\k{a}\.{z}ewski}},\ }\href@noop {} {\bibfield  {journal} {\bibinfo
  {journal} {Physical Review A},\ }\textbf {\bibinfo {volume} {65}},\ \bibinfo
  {pages} {063605} (\bibinfo {year} {2002})}\BibitemShut {NoStop}%
\bibitem [{\citenamefont {Ziń}\ \emph {et~al.}(2005)\citenamefont {Ziń},
  \citenamefont {Chwedeńczuk}, \citenamefont {Veitia}, \citenamefont
  {Rz\k{a}\.{z}ewski},\ and\ \citenamefont {Trippenbach}}]{Zin2005}%
  \BibitemOpen
  \bibfield  {author} {\bibinfo {author} {\bibfnamefont {P.}~\bibnamefont
  {Ziń}}, \bibinfo {author} {\bibfnamefont {J.}~\bibnamefont {Chwedeńczuk}},
  \bibinfo {author} {\bibfnamefont {A.}~\bibnamefont {Veitia}}, \bibinfo
  {author} {\bibfnamefont {K.}~\bibnamefont {Rz\k{a}\.{z}ewski}}, \ and\
  \bibinfo {author} {\bibfnamefont {M.}~\bibnamefont {Trippenbach}},\
  }\href@noop {} {\bibfield  {journal} {\bibinfo  {journal} {Physical Review
  Letters},\ }\textbf {\bibinfo {volume} {94}},\ \bibinfo {pages} {200401}
  (\bibinfo {year} {2005})}\BibitemShut {NoStop}%
\bibitem [{\citenamefont {Ziń}\ \emph {et~al.}(2006)\citenamefont {Ziń},
  \citenamefont {Chwedeńczuk},\ and\ \citenamefont {Trippenbach}}]{Zin2006}%
  \BibitemOpen
  \bibfield  {author} {\bibinfo {author} {\bibfnamefont {P.}~\bibnamefont
  {Ziń}}, \bibinfo {author} {\bibfnamefont {J.}~\bibnamefont {Chwedeńczuk}},
  \ and\ \bibinfo {author} {\bibfnamefont {M.}~\bibnamefont {Trippenbach}},\
  }\href@noop {} {\bibfield  {journal} {\bibinfo  {journal} {Physical Review
  A},\ }\textbf {\bibinfo {volume} {73}},\ \bibinfo {pages} {033602} (\bibinfo
  {year} {2006})}\BibitemShut {NoStop}%
\bibitem [{\citenamefont {Kheruntsyan}\ and\ \citenamefont
  {Drummond}(2002)}]{Kheruntsyan2002}%
  \BibitemOpen
  \bibfield  {author} {\bibinfo {author} {\bibfnamefont {K.~V.}\ \bibnamefont
  {Kheruntsyan}}\ and\ \bibinfo {author} {\bibfnamefont {P.}~\bibnamefont
  {Drummond}},\ }\href@noop {} {\bibfield  {journal} {\bibinfo  {journal}
  {Physical Review A},\ }\textbf {\bibinfo {volume} {66}},\ \bibinfo {pages}
  {031602} (\bibinfo {year} {2002})}\BibitemShut {NoStop}%
\bibitem [{\citenamefont {Kheruntsyan}\ \emph {et~al.}(2005)\citenamefont
  {Kheruntsyan}, \citenamefont {Olsen},\ and\ \citenamefont
  {Drummond}}]{Kheruntsyan2005a}%
  \BibitemOpen
  \bibfield  {author} {\bibinfo {author} {\bibfnamefont {K.~V.}\ \bibnamefont
  {Kheruntsyan}}, \bibinfo {author} {\bibfnamefont {M.~K.}\ \bibnamefont
  {Olsen}}, \ and\ \bibinfo {author} {\bibfnamefont {P.~D.}\ \bibnamefont
  {Drummond}},\ }\href@noop {} {\bibfield  {journal} {\bibinfo  {journal}
  {Physical Review Letters},\ }\textbf {\bibinfo {volume} {95}},\ \bibinfo
  {pages} {150405} (\bibinfo {year} {2005})}\BibitemShut {NoStop}%
\bibitem [{\citenamefont {Kheruntsyan}(2005)}]{Kheruntsyan2005b}%
  \BibitemOpen
  \bibfield  {author} {\bibinfo {author} {\bibfnamefont {K.~V.}\ \bibnamefont
  {Kheruntsyan}},\ }\href@noop {} {\bibfield  {journal} {\bibinfo  {journal}
  {Physical Review A},\ }\textbf {\bibinfo {volume} {71}},\ \bibinfo {pages}
  {053609} (\bibinfo {year} {2005})}\BibitemShut {NoStop}%
\bibitem [{\citenamefont {Deuar}\ and\ \citenamefont
  {Drummond}(2007)}]{Deuar2007}%
  \BibitemOpen
  \bibfield  {author} {\bibinfo {author} {\bibfnamefont {P.}~\bibnamefont
  {Deuar}}\ and\ \bibinfo {author} {\bibfnamefont {P.}~\bibnamefont
  {Drummond}},\ }\href@noop {} {\bibfield  {journal} {\bibinfo  {journal}
  {Physical Review Letters},\ }\textbf {\bibinfo {volume} {98}},\ \bibinfo
  {pages} {120402} (\bibinfo {year} {2007})}\BibitemShut {NoStop}%
\bibitem [{\citenamefont {Perrin}\ \emph {et~al.}(2008)\citenamefont {Perrin},
  \citenamefont {Savage}, \citenamefont {Boiron}, \citenamefont
  {Krachmalnicoff}, \citenamefont {Westbrook},\ and\ \citenamefont
  {Kheruntsyan}}]{Perrin2008}%
  \BibitemOpen
  \bibfield  {author} {\bibinfo {author} {\bibfnamefont {A.}~\bibnamefont
  {Perrin}}, \bibinfo {author} {\bibfnamefont {C.~M.}\ \bibnamefont {Savage}},
  \bibinfo {author} {\bibfnamefont {D.}~\bibnamefont {Boiron}}, \bibinfo
  {author} {\bibfnamefont {V.}~\bibnamefont {Krachmalnicoff}}, \bibinfo
  {author} {\bibfnamefont {C.~I.}\ \bibnamefont {Westbrook}}, \ and\ \bibinfo
  {author} {\bibfnamefont {K.~V.}\ \bibnamefont {Kheruntsyan}},\ }\href@noop {}
  {\bibfield  {journal} {\bibinfo  {journal} {New Journal of Physics},\
  }\textbf {\bibinfo {volume} {10}},\ \bibinfo {pages} {045021} (\bibinfo
  {year} {2008})}\BibitemShut {NoStop}%
\bibitem [{\citenamefont {Carusotto}\ \emph {et~al.}(2001)\citenamefont
  {Carusotto}, \citenamefont {Castin},\ and\ \citenamefont
  {Dalibard}}]{Carusotto2001}%
  \BibitemOpen
  \bibfield  {author} {\bibinfo {author} {\bibfnamefont {I.}~\bibnamefont
  {Carusotto}}, \bibinfo {author} {\bibfnamefont {Y.}~\bibnamefont {Castin}}, \
  and\ \bibinfo {author} {\bibfnamefont {J.}~\bibnamefont {Dalibard}},\
  }\href@noop {} {\bibfield  {journal} {\bibinfo  {journal} {Physical Review
  A},\ }\textbf {\bibinfo {volume} {63}},\ \bibinfo {pages} {023606} (\bibinfo
  {year} {2001})}\BibitemShut {NoStop}%
\bibitem [{\citenamefont {Norrie}\ \emph {et~al.}(2005)\citenamefont {Norrie},
  \citenamefont {Ballagh},\ and\ \citenamefont {Gardiner}}]{Norrie2005}%
  \BibitemOpen
  \bibfield  {author} {\bibinfo {author} {\bibfnamefont {A.}~\bibnamefont
  {Norrie}}, \bibinfo {author} {\bibfnamefont {R.}~\bibnamefont {Ballagh}}, \
  and\ \bibinfo {author} {\bibfnamefont {C.}~\bibnamefont {Gardiner}},\
  }\href@noop {} {\bibfield  {journal} {\bibinfo  {journal} {Physical Review
  Letters},\ }\textbf {\bibinfo {volume} {94}},\ \bibinfo {pages} {040401}
  (\bibinfo {year} {2005})}\BibitemShut {NoStop}%
\bibitem [{\citenamefont {Norrie}\ \emph {et~al.}(2006)\citenamefont {Norrie},
  \citenamefont {Ballagh},\ and\ \citenamefont {Gardiner}}]{Norrie2006}%
  \BibitemOpen
  \bibfield  {author} {\bibinfo {author} {\bibfnamefont {A.}~\bibnamefont
  {Norrie}}, \bibinfo {author} {\bibfnamefont {R.}~\bibnamefont {Ballagh}}, \
  and\ \bibinfo {author} {\bibfnamefont {C.}~\bibnamefont {Gardiner}},\
  }\href@noop {} {\bibfield  {journal} {\bibinfo  {journal} {Physical Review
  A},\ }\textbf {\bibinfo {volume} {73}},\ \bibinfo {pages} {043617} (\bibinfo
  {year} {2006})}\BibitemShut {NoStop}%
\bibitem [{\citenamefont {Spielman}\ \emph {et~al.}(2006)\citenamefont
  {Spielman}, \citenamefont {Johnson}, \citenamefont {Huckans}, \citenamefont
  {Fertig}, \citenamefont {Rolston}, \citenamefont {Phillips},\ and\
  \citenamefont {Porto}}]{Spielman2006}%
  \BibitemOpen
  \bibfield  {author} {\bibinfo {author} {\bibfnamefont {I.~B.}\ \bibnamefont
  {Spielman}}, \bibinfo {author} {\bibfnamefont {P.~R.}\ \bibnamefont
  {Johnson}}, \bibinfo {author} {\bibfnamefont {J.~H.}\ \bibnamefont
  {Huckans}}, \bibinfo {author} {\bibfnamefont {C.~D.}\ \bibnamefont {Fertig}},
  \bibinfo {author} {\bibfnamefont {S.~L.}\ \bibnamefont {Rolston}}, \bibinfo
  {author} {\bibfnamefont {W.~D.}\ \bibnamefont {Phillips}}, \ and\ \bibinfo
  {author} {\bibfnamefont {J.~V.}\ \bibnamefont {Porto}},\ }\href@noop {}
  {\bibfield  {journal} {\bibinfo  {journal} {Physical Review A},\ }\textbf
  {\bibinfo {volume} {73}},\ \bibinfo {pages} {020702} (\bibinfo {year}
  {2006})}\BibitemShut {NoStop}%
\bibitem [{\citenamefont {Krachmalnicoff}\ \emph {et~al.}(2010)\citenamefont
  {Krachmalnicoff}, \citenamefont {Jaskula}, \citenamefont {Bonneau},
  \citenamefont {Leung}, \citenamefont {Partridge}, \citenamefont {Boiron},
  \citenamefont {Westbrook}, \citenamefont {Deuar}, \citenamefont {Zin},
  \citenamefont {Trippenbach},\ and\ \citenamefont
  {Kheruntsyan}}]{Krachmalnicoff2010}%
  \BibitemOpen
  \bibfield  {author} {\bibinfo {author} {\bibfnamefont {V.}~\bibnamefont
  {Krachmalnicoff}}, \bibinfo {author} {\bibfnamefont {J.-C.}\ \bibnamefont
  {Jaskula}}, \bibinfo {author} {\bibfnamefont {M.}~\bibnamefont {Bonneau}},
  \bibinfo {author} {\bibfnamefont {V.}~\bibnamefont {Leung}}, \bibinfo
  {author} {\bibfnamefont {G.~B.}\ \bibnamefont {Partridge}}, \bibinfo {author}
  {\bibfnamefont {D.}~\bibnamefont {Boiron}}, \bibinfo {author} {\bibfnamefont
  {C.~I.}\ \bibnamefont {Westbrook}}, \bibinfo {author} {\bibfnamefont
  {P.}~\bibnamefont {Deuar}}, \bibinfo {author} {\bibfnamefont
  {P.}~\bibnamefont {Zin}}, \bibinfo {author} {\bibfnamefont {M.}~\bibnamefont
  {Trippenbach}}, \ and\ \bibinfo {author} {\bibfnamefont {K.~V.}\ \bibnamefont
  {Kheruntsyan}},\ }\href@noop {} {\bibfield  {journal} {\bibinfo  {journal}
  {Physical Review Letters},\ }\textbf {\bibinfo {volume} {104}},\ \bibinfo
  {pages} {150402} (\bibinfo {year} {2010})}\BibitemShut {NoStop}%
\bibitem [{\citenamefont {Javanainen}\ and\ \citenamefont
  {Ivanov}(1999)}]{javanainen:99}%
  \BibitemOpen
  \bibfield  {author} {\bibinfo {author} {\bibfnamefont {J.}~\bibnamefont
  {Javanainen}}\ and\ \bibinfo {author} {\bibfnamefont {M.~Y.}\ \bibnamefont
  {Ivanov}},\ }\href@noop {} {\bibfield  {journal} {\bibinfo  {journal}
  {Physical Review A},\ }\textbf {\bibinfo {volume} {60}},\ \bibinfo {pages}
  {2351} (\bibinfo {year} {1999})}\BibitemShut {NoStop}%
\bibitem [{\citenamefont {B\"{u}cker}\ \emph {et~al.}(2009)\citenamefont
  {B\"{u}cker}, \citenamefont {Perrin}, \citenamefont {Manz}, \citenamefont
  {Betz}, \citenamefont {Koller}, \citenamefont {Plisson}, \citenamefont
  {Rottmann}, \citenamefont {Schumm},\ and\ \citenamefont
  {Schmiedmayer}}]{Buecker2009}%
  \BibitemOpen
  \bibfield  {author} {\bibinfo {author} {\bibfnamefont {R.}~\bibnamefont
  {B\"{u}cker}}, \bibinfo {author} {\bibfnamefont {A.}~\bibnamefont {Perrin}},
  \bibinfo {author} {\bibfnamefont {S.}~\bibnamefont {Manz}}, \bibinfo {author}
  {\bibfnamefont {T.}~\bibnamefont {Betz}}, \bibinfo {author} {\bibfnamefont
  {C.}~\bibnamefont {Koller}}, \bibinfo {author} {\bibfnamefont
  {T.}~\bibnamefont {Plisson}}, \bibinfo {author} {\bibfnamefont
  {J.}~\bibnamefont {Rottmann}}, \bibinfo {author} {\bibfnamefont
  {T.}~\bibnamefont {Schumm}}, \ and\ \bibinfo {author} {\bibfnamefont
  {J.}~\bibnamefont {Schmiedmayer}},\ }\href@noop {} {\bibfield  {journal}
  {\bibinfo  {journal} {New Journal of Physics},\ }\textbf {\bibinfo {volume}
  {11}},\ \bibinfo {pages} {103039} (\bibinfo {year} {2009})}\BibitemShut
  {NoStop}%
\bibitem [{\citenamefont {Reichel}\ and\ \citenamefont
  {Vuletic}(2011)}]{Reichel2011}%
  \BibitemOpen
  \bibinfo {editor} {\bibfnamefont {J.}~\bibnamefont {Reichel}}\ and\ \bibinfo
  {editor} {\bibfnamefont {V.}~\bibnamefont {Vuletic}},\ eds.,\ \href@noop {}
  {\emph {\bibinfo {title} {{Atom Chips}}}}\ (\bibinfo  {publisher}
  {Wiley-VCH},\ \bibinfo {year} {2011})\ ISBN \bibinfo {isbn}
  {3-527-40755-3}\BibitemShut {NoStop}%
\bibitem [{\citenamefont {Trinker}\ \emph {et~al.}(2008)\citenamefont
  {Trinker}, \citenamefont {Groth}, \citenamefont {Haslinger}, \citenamefont
  {Manz}, \citenamefont {Betz}, \citenamefont {Schneider}, \citenamefont
  {Bar-Joseph}, \citenamefont {Schumm},\ and\ \citenamefont
  {Schmiedmayer}}]{Trinker2008a}%
  \BibitemOpen
  \bibfield  {author} {\bibinfo {author} {\bibfnamefont {M.}~\bibnamefont
  {Trinker}}, \bibinfo {author} {\bibfnamefont {S.}~\bibnamefont {Groth}},
  \bibinfo {author} {\bibfnamefont {S.}~\bibnamefont {Haslinger}}, \bibinfo
  {author} {\bibfnamefont {S.}~\bibnamefont {Manz}}, \bibinfo {author}
  {\bibfnamefont {T.}~\bibnamefont {Betz}}, \bibinfo {author} {\bibfnamefont
  {S.}~\bibnamefont {Schneider}}, \bibinfo {author} {\bibfnamefont
  {I.}~\bibnamefont {Bar-Joseph}}, \bibinfo {author} {\bibfnamefont
  {T.}~\bibnamefont {Schumm}}, \ and\ \bibinfo {author} {\bibfnamefont
  {J.}~\bibnamefont {Schmiedmayer}},\ }\href@noop {} {\bibfield  {journal}
  {\bibinfo  {journal} {Applied Physics Letters},\ }\textbf {\bibinfo {volume}
  {92}},\ \bibinfo {pages} {254102} (\bibinfo {year} {2008})}\BibitemShut
  {NoStop}%
\bibitem [{Note1()}]{Note1}%
  \BibitemOpen
  \bibinfo {note} {Note that this procedure is not equivalent to using
  numerically optimized excitations for different efficiencies, and may lead
  e.g. to increased collective oscillations after the excitation, which is
  irrelevant for the emission dynamics.}\BibitemShut {Stop}%
\bibitem [{\citenamefont {de~Oliveira}\ \emph {et~al.}(1990)\citenamefont
  {de~Oliveira}, \citenamefont {Kim}, \citenamefont {Knight},\ and\
  \citenamefont {Buek}}]{DeOliveira1990}%
  \BibitemOpen
  \bibfield  {author} {\bibinfo {author} {\bibfnamefont {F.}~\bibnamefont
  {de~Oliveira}}, \bibinfo {author} {\bibfnamefont {M.}~\bibnamefont {Kim}},
  \bibinfo {author} {\bibfnamefont {P.}~\bibnamefont {Knight}}, \ and\ \bibinfo
  {author} {\bibfnamefont {V.}~\bibnamefont {Buek}},\ }\href@noop {} {\bibfield
   {journal} {\bibinfo  {journal} {Physical Review A},\ }\textbf {\bibinfo
  {volume} {41}},\ \bibinfo {pages} {2645} (\bibinfo {year}
  {1990})}\BibitemShut {NoStop}%
\bibitem [{\citenamefont {Petrov}\ \emph {et~al.}(2000)\citenamefont {Petrov},
  \citenamefont {Shlyapnikov},\ and\ \citenamefont {Walraven}}]{petrov:00}%
  \BibitemOpen
  \bibfield  {author} {\bibinfo {author} {\bibfnamefont {D.~S.}\ \bibnamefont
  {Petrov}}, \bibinfo {author} {\bibfnamefont {G.~V.}\ \bibnamefont
  {Shlyapnikov}}, \ and\ \bibinfo {author} {\bibfnamefont {J.~T.~M.}\
  \bibnamefont {Walraven}},\ }\href@noop {} {\bibfield  {journal} {\bibinfo
  {journal} {Physical Review Letters},\ }\textbf {\bibinfo {volume} {85}},\
  \bibinfo {pages} {3745} (\bibinfo {year} {2000})}\BibitemShut {NoStop}%
\bibitem [{Note2()}]{Note2}%
  \BibitemOpen
  \bibinfo {note} {Note that the non-condensed modes, described by $\rho
  _{th}(x,x')$ in eq.~(\ref {eq:condensate}), rapidly re-thermalize~\cite
  {Mazets2011}, impeding an unambiguous determination of $T$ from experimental
  data. The temperature-dependent emission and thermalization dynamics will be
  subject of a future publication.}\BibitemShut {Stop}%
\bibitem [{\citenamefont {Kheruntsyan}\ \emph {et~al.}(2012)\citenamefont
  {Kheruntsyan}, \citenamefont {Jaskula}, \citenamefont {Deuar}, \citenamefont
  {Bonneau}, \citenamefont {Partridge}, \citenamefont {Ruaudel}, \citenamefont
  {Lopes}, \citenamefont {Boiron},\ and\ \citenamefont
  {Westbrook}}]{Kheruntsyan2012}%
  \BibitemOpen
  \bibfield  {author} {\bibinfo {author} {\bibfnamefont {K.~V.}\ \bibnamefont
  {Kheruntsyan}}, \bibinfo {author} {\bibfnamefont {J.~C.}\ \bibnamefont
  {Jaskula}}, \bibinfo {author} {\bibfnamefont {P.}~\bibnamefont {Deuar}},
  \bibinfo {author} {\bibfnamefont {M.}~\bibnamefont {Bonneau}}, \bibinfo
  {author} {\bibfnamefont {G.~B.}\ \bibnamefont {Partridge}}, \bibinfo {author}
  {\bibfnamefont {J.}~\bibnamefont {Ruaudel}}, \bibinfo {author} {\bibfnamefont
  {R.}~\bibnamefont {Lopes}}, \bibinfo {author} {\bibfnamefont
  {D.}~\bibnamefont {Boiron}}, \ and\ \bibinfo {author} {\bibfnamefont {C.~I.}\
  \bibnamefont {Westbrook}},\ }\href@noop {} { (\bibinfo {year} {2012})},\
  \Eprint {http://arxiv.org/abs/1204.0058} {arXiv:1204.0058} \BibitemShut
  {NoStop}%
\bibitem [{\citenamefont {Lesanovsky}\ \emph {et~al.}(2006)\citenamefont
  {Lesanovsky}, \citenamefont {Schumm}, \citenamefont {Hofferberth},
  \citenamefont {Andersson}, \citenamefont {Kr\"{u}ger},\ and\ \citenamefont
  {Schmiedmayer}}]{Lesanovsky2006}%
  \BibitemOpen
  \bibfield  {author} {\bibinfo {author} {\bibfnamefont {I.}~\bibnamefont
  {Lesanovsky}}, \bibinfo {author} {\bibfnamefont {T.}~\bibnamefont {Schumm}},
  \bibinfo {author} {\bibfnamefont {S.}~\bibnamefont {Hofferberth}}, \bibinfo
  {author} {\bibfnamefont {L.~M.}\ \bibnamefont {Andersson}}, \bibinfo {author}
  {\bibfnamefont {P.}~\bibnamefont {Kr\"{u}ger}}, \ and\ \bibinfo {author}
  {\bibfnamefont {J.}~\bibnamefont {Schmiedmayer}},\ }\href@noop {} {\bibfield
  {journal} {\bibinfo  {journal} {Physical Review A},\ }\textbf {\bibinfo
  {volume} {73}},\ \bibinfo {pages} {033619} (\bibinfo {year}
  {2006})}\BibitemShut {NoStop}%
\bibitem [{\citenamefont {Schumm}\ \emph {et~al.}(2005)\citenamefont {Schumm},
  \citenamefont {Hofferberth}, \citenamefont {Andersson}, \citenamefont
  {Wildermuth}, \citenamefont {Groth}, \citenamefont {Bar-Joseph},
  \citenamefont {Schmiedmayer},\ and\ \citenamefont
  {Kr\"{u}ger}}]{Schumm2005a}%
  \BibitemOpen
  \bibfield  {author} {\bibinfo {author} {\bibfnamefont {T.}~\bibnamefont
  {Schumm}}, \bibinfo {author} {\bibfnamefont {S.}~\bibnamefont {Hofferberth}},
  \bibinfo {author} {\bibfnamefont {L.~M.}\ \bibnamefont {Andersson}}, \bibinfo
  {author} {\bibfnamefont {S.}~\bibnamefont {Wildermuth}}, \bibinfo {author}
  {\bibfnamefont {S.}~\bibnamefont {Groth}}, \bibinfo {author} {\bibfnamefont
  {I.}~\bibnamefont {Bar-Joseph}}, \bibinfo {author} {\bibfnamefont
  {J.}~\bibnamefont {Schmiedmayer}}, \ and\ \bibinfo {author} {\bibfnamefont
  {P.}~\bibnamefont {Kr\"{u}ger}},\ }\href@noop {} {\bibfield  {journal}
  {\bibinfo  {journal} {Nature Physics},\ }\textbf {\bibinfo {volume} {1}},\
  \bibinfo {pages} {57} (\bibinfo {year} {2005})}\BibitemShut {NoStop}%
\bibitem [{\citenamefont {Shirley}(1965)}]{Shirley1965}%
  \BibitemOpen
  \bibfield  {author} {\bibinfo {author} {\bibfnamefont {J.~H.}\ \bibnamefont
  {Shirley}},\ }\href@noop {} {\bibfield  {journal} {\bibinfo  {journal}
  {Physical Review},\ }\textbf {\bibinfo {volume} {138}},\ \bibinfo {pages}
  {B979} (\bibinfo {year} {1965})}\BibitemShut {NoStop}%
\bibitem [{\citenamefont {Jacqmin}\ \emph {et~al.}()\citenamefont {Jacqmin}
  \emph {et~al.}}]{Jacqmin2012}%
  \BibitemOpen
  \bibfield  {author} {\bibinfo {author} {\bibfnamefont {T.}~\bibnamefont
  {Jacqmin}} \emph {et~al.},\ }\href@noop {} {}\bibinfo {note} {In
  preparation}\BibitemShut {NoStop}%
\bibitem [{\citenamefont {Jirari}\ \emph {et~al.}(2009)\citenamefont {Jirari},
  \citenamefont {Hekking},\ and\ \citenamefont {Buisson}}]{Jirari2009}%
  \BibitemOpen
  \bibfield  {author} {\bibinfo {author} {\bibfnamefont {H.}~\bibnamefont
  {Jirari}}, \bibinfo {author} {\bibfnamefont {F.~W.~J.}\ \bibnamefont
  {Hekking}}, \ and\ \bibinfo {author} {\bibfnamefont {O.}~\bibnamefont
  {Buisson}},\ }\href@noop {} {\bibfield  {journal} {\bibinfo  {journal} {EPL
  (Europhysics Letters)},\ }\textbf {\bibinfo {volume} {87}},\ \bibinfo {pages}
  {28004} (\bibinfo {year} {2009})}\BibitemShut {NoStop}%
\bibitem [{\citenamefont {Khani}\ \emph {et~al.}(2009)\citenamefont {Khani},
  \citenamefont {Gambetta}, \citenamefont {Motzoi},\ and\ \citenamefont
  {Wilhelm}}]{Khani2009}%
  \BibitemOpen
  \bibfield  {author} {\bibinfo {author} {\bibfnamefont {B.}~\bibnamefont
  {Khani}}, \bibinfo {author} {\bibfnamefont {J.~M.}\ \bibnamefont {Gambetta}},
  \bibinfo {author} {\bibfnamefont {F.}~\bibnamefont {Motzoi}}, \ and\ \bibinfo
  {author} {\bibfnamefont {F.~K.}\ \bibnamefont {Wilhelm}},\ }\href@noop {}
  {\bibfield  {journal} {\bibinfo  {journal} {Physica Scripta},\ }\textbf
  {\bibinfo {volume} {T137}},\ \bibinfo {pages} {014021} (\bibinfo {year}
  {2009})}\BibitemShut {NoStop}%
\bibitem [{\citenamefont {Alon}\ \emph {et~al.}(2008)\citenamefont {Alon},
  \citenamefont {Streltsov},\ and\ \citenamefont {Cederbaum}}]{alon:08}%
  \BibitemOpen
  \bibfield  {author} {\bibinfo {author} {\bibfnamefont {O.~E.}\ \bibnamefont
  {Alon}}, \bibinfo {author} {\bibfnamefont {A.~I.}\ \bibnamefont {Streltsov}},
  \ and\ \bibinfo {author} {\bibfnamefont {L.~S.}\ \bibnamefont {Cederbaum}},\
  }\href@noop {} {\bibfield  {journal} {\bibinfo  {journal} {Physical Review
  A},\ }\textbf {\bibinfo {volume} {77}},\ \bibinfo {pages} {033613} (\bibinfo
  {year} {2008})}\BibitemShut {NoStop}%
\bibitem [{\citenamefont {Grond}\ \emph {et~al.}(2009)\citenamefont {Grond},
  \citenamefont {von Winckel}, \citenamefont {Schmiedmayer},\ and\
  \citenamefont {Hohenester}}]{grond.pra:09b}%
  \BibitemOpen
  \bibfield  {author} {\bibinfo {author} {\bibfnamefont {J.}~\bibnamefont
  {Grond}}, \bibinfo {author} {\bibfnamefont {G.}~\bibnamefont {von Winckel}},
  \bibinfo {author} {\bibfnamefont {J.}~\bibnamefont {Schmiedmayer}}, \ and\
  \bibinfo {author} {\bibfnamefont {U.}~\bibnamefont {Hohenester}},\
  }\href@noop {} {\bibfield  {journal} {\bibinfo  {journal} {Physical Review
  A},\ }\textbf {\bibinfo {volume} {80}},\ \bibinfo {pages} {053625} (\bibinfo
  {year} {2009})}\BibitemShut {NoStop}%
\bibitem [{\citenamefont {Mazets}(2011)}]{Mazets2011}%
  \BibitemOpen
  \bibfield  {author} {\bibinfo {author} {\bibfnamefont {I.}~\bibnamefont
  {Mazets}},\ }\href@noop {} {\bibfield  {journal} {\bibinfo  {journal}
  {Physical Review A},\ }\textbf {\bibinfo {volume} {83}},\ \bibinfo {pages}
  {043625} (\bibinfo {year} {2011})}\BibitemShut {NoStop}%
\end{thebibliography}
%

\end{document}